\documentclass[11pt]{article}
\def\ben{\begin{enumerate}} \def\een{\end{enumerate}}
\def\beq{\begin{equation}} \def\eeq{\end{equation}}
\def\bea{\begin{eqnarray}} \def\eea{\end{eqnarray}}
\def\beann{\begin{eqnarray*}} \def\eeann{\end{eqnarray*}}
\def\beasn{\begin{sneqnarray}} \def\eeasn{\end{sneqnarray}}
\newcommand{\bref}[1]{(\ref{#1})}

\begin{document}

\title{\bf Fluctuations around classical solutions for gauge theories
 in Lagrangian and Hamiltonian approach}


\author{Olivera Mi\v{s}kovi\'c $^a$ and Josep M. Pons $^b $ \\
\small\tt{olivera@fis.puc.cl\,,\qquad pons@ecm.ub.es\qquad\qquad}\\
$^a$\small\emph{Departamento de F\'\i sica, P. Universidad Cat\'olica de
Chile, Casilla 306, Santiago 22, Chile.}\\
$^b$\small\emph{Departament d'Estructura i Constituents de
la Mat\`eria, Facultat de F\'\i sica,
}\\
\small\emph{Universitat de Barcelona, Av.~Diagonal 647,
08028 Barcelona, Catalonia, Spain.}}
\date{}
\maketitle

\begin{abstract}

We analyze the dynamics of gauge theories and constrained systems
in general under small perturbations around a classical solution
in both Lagrangian and Hamiltonian formalisms. We
prove that a fluctuations theory, described by a quadratic
Lagrangian, has the same constraint structure and number of
physical degrees of freedom as the original non-perturbed theory,
assuming the non-degenerate solution has been chosen. We show that
the number of Noether gauge symmetries is the same in both
theories, but that the gauge algebra in the fluctuations theory
becomes Abelianized. We also show that the fluctuations theory
inherits all functionally independent rigid symmetries from the
original theory, and that these symmetries are generated by linear
or quadratic generators according to whether the original symmetry
is preserved by the background, or is broken by it. We illustrate
these results with the examples.

\end{abstract}

\section{Introduction }
\label{sec:intro}

Dynamics of linearized perturbations, obeying the equations of
motion of the quadratic action formulated around a classical
solution (background) of a field theory, has been widely used for
numerous applications.\footnote{We use indistinctly the words
perturbations and fluctuations, the latter one corresponding more
closely to the language usual in quantum theory and statistical
mechanics.} Thus, it is used as a test of stability, where the
fluctuations around a stable solution ---i.e., vacuum--- have
harmonic oscillator dynamics. In general, the quadratic potentials
also provide quantum corrections for an \textquotedblleft
effective\textquotedblright\ mass of a solution, which can be a
wave packet such as a soliton, or can identify tachyonic modes
(with negative square mass) which would signal an instability, and
so on. Let us emphasize that here we perturb only fundamental
fields in a theory, without making any expansion in the coupling
constant, nor the quantum loop expansion in $\hbar $. These, and
many other numerous uses of the fluctuations theory, such as
quantization of solitons, spontaneous breaking of symmetry,
gravitational waves, etc., have become so standard in physics that
they can be found in any textbook on field theory.

Intuitively, the dynamics of the original and \textquotedblleft
linearized" (described by a quadratic action) theories should
exhibit some parallelisms. In gauge systems, however, the presence
of unphysical degrees of freedom, and the frequent appearance of
constraints, obscures this intuition. For example, there are
systems which seem to have more degrees of freedom when linearized
around some backgrounds \cite{Chandia-Troncoso-Zanelli}. In these
quadratic theories, the gauge symmetry appears as broken with
respect to the full theory.

In addition, and connected with the previous observation, one may
ask what is a criterion for the Legendre transformation to commute
with the process of getting the theory of quadratic fluctuations
in both Lagrangian and Hamiltonian approaches.

Other questions that may arise are whether the quadratic action
contains the same rigid and gauge symmetries as the original one,
how its constraint structure looks like and how the corresponding
canonical theory is formulated.

In this paper we address all these questions. We write out a
criterion that guarantees that the quadratic action contains as
much gauge freedom as the original one. In fact we show that the
gauge algebra, if it was originally non-Abelian, in the
fluctuations theory becomes Abelianized. This agrees with the fact
that a non-Abelian theory cannot be described by a quadratic
action, but it requires higher-order terms.

We also show that the canonical quadratic Hamiltonian for the
fluctuations is built up from two different pieces of information:
one is obviously the quadratic term of the expansion of the
original canonical Hamiltonian, whereas the other, not so obvious,
consists in the quadratic terms of the expansion of the original
primary constraints. This result is quite natural from the
viewpoint of the Dirac-Bergman theory of constrained systems, on
which we rely throughout the paper. In connecting the Lagrangian
and the Hamiltonian formulations at the level of the original
action with those at the level of the quadratic action, we see
that a mismatch appears between the respective Legendre maps (from
tangent space to phase space), but that such mismatch is of higher
order in the fluctuations and thus does not affect the consistency
of our procedure.

Concerning the constraint algorithm in the fluctuations theory
---either in the Lagrangian, or in the Hamiltonian formulation---
it is shown that it reflects the structure of the algorithm that
holds for the original theory.  This result is not a
priori obvious, because when there is more than one generation of
constraints, that is, when new constraints arise from evolution of
original primary constraints, then the process of truncation of
higher order terms may not commute with taking the time derivative
and Poisson bracket. In particular, we show that the original
Second Class constraints yield Second Class constraints for the
fluctuations theory, and First Class constraints yield {\sl
Abelianized} First Class constraints.

Noether symmetries and conserved quantities are also shown to be
inherited from the original theory to the fluctuations theory, but
in a non-straightforward way. In fact, due to the presence of the
classical solution --the background-- the original Noether
symmetries split according to whether they are preserved by the
background, or are broken by it. Remarkably, it turns out that
those that are respected by the background yield rigid symmetries
for the quadratic action with {\sl quadratic} generators, whereas
the broken symmetries yield symmetries with {\sl linear}
generators. The importance of quadratic generators is noteworthy
in field theories with supersymetry where a BPS state, a solution
which preserves some supersymmetries, is preferred as a ground
state since it plays significant role in the stability of a
theory.

Regarding quantization around the classical solution, note that
the gauge fixing is technically simpler for the quadratic theory
than for the original theory, owing to the Abelian structure of
the new gauge group. This means that one could have spared the
technicalities of gauge fixing in the non-Abelian case and proceed
instead to the easier gauge fixing for the quadratic --Abelian--
theory around the classical solution.

After introducing some notation in the next section, we address
the tangent space version and the canonical version of the
fluctuations theory in Sections 3 and 4, where the connection
between both formalisms is analyzed as well as their constraint
algorithms. In Section 5 we study the Noether symmetries for the
fluctuations theory. Examples are discussed in Section 6, and
Section 7 is devoted to conclusions.

\section{Notation}
\label{sec:nut}

We will use for simplicity the language of mechanics. Since our
prime interest are gauge field theories, a quick switch to the
field theory language can be achieved by using DeWitt's condensed
notation \cite{DeWitt}.

Consider the dynamics of a classical mechanical system with finite
number of degrees of freedom, described by a Lagrangian
$L(q,\,\dot q)$ depending at most on first derivatives, up to
divergence terms, and which does not depend on time explicitly
(first-order systems). The local coordinates $q^i$ ($i=1,\ldots
,n$) parameterize a configuration manifold $\cal Q$ of dimension
$n$, and therefore the entire dynamics of the system happens on
the corresponding tangent bundle $T\mathcal Q$ which is a
configuration-velocity space $(q,\,\dot q)$. The Euler-Lagrange
equations of motion (e.o.m.) are\footnote{All functions are
assumed to be continuous and differentiable as many times as the
formalism requires.}
$$
[L]_i := \alpha_i - W_{ij}\,\ddot q^j=0\,,
$$
with the Hessian matrix
\begin{equation}
W_{ij}\equiv {\partial^2L\over\partial\dot
q^i\partial\dot q^j}\,,
\label{hess}
\end{equation}
and
$$
\alpha_i :=
    - {\partial^2L\over\partial\dot q^i\partial q^j}\,\dot q^j
    + {\partial L\over\partial q^i} \,.
$$
Regular systems have invertible Hessian. We are, however,
interested in singular Lagrangians with non-invertible Hessian
matrices, also called constrained systems
\cite{Dirac1950}--\cite{Anderson-Bergmann}, since the gauge
theories rely on them. \vspace{4mm}

In order to pass to the Hamiltonian formalism, we apply
the Legendre map ${\mathcal F}\!L: \ T{\mathcal Q} \rightarrow T^*{\mathcal Q}$
to the original theory, which maps configurational space into
phase space,
$$(q,\dot q) \rightarrow (q,p=\hat p(q,\dot q))\,,$$
where the momentum map is
$$
\hat p(q,\dot q):=\frac{\partial L}{\partial \dot q}\,.
$$
We made the assumption that the rank of the Hessian matrix is
constant everywhere. If this condition is not satisfied throughout
the whole tangent bundle, we restrict our considerations to a
region of it, with the same dimensionality, where this condition
holds. For degenerate systems with non-constant rank of $W$, see
\cite{Saavedra-Troncoso-Zanelli}. So we are assuming that the rank
of the Legendre map ${\mathcal F}\!L$ from the tangent bundle
$T{\mathcal Q}$ to the cotangent bundle $T^*{\mathcal Q}$ is
constant throughout $T{\mathcal Q}$ and equal to, say, $2n-k$. The
image of ${\mathcal F}\!L$ is locally defined by the vanishing of
$k$ independent functions, $\phi_\mu(q, p), \mu = 1,2,..,k$. These
functions are the {\sl primary constraints}, and their pullback
${\mathcal F}\!L^*\phi_\mu$ to the tangent bundle is identically
zero:
\begin{equation}({\mathcal F}\!L^*\phi_\mu)(q,\dot q):=\phi_\mu(q, \hat p)
=0\,,\ \ \forall\, q,\dot q\,. \label{pullback}
\end{equation}

The primary constraints form a generating set of the ideal of
functions that vanish on the image of the Legendre map. With their
help it is easy to obtain a basis of null vectors for the Hessian
matrix \cite{Batlle-Gomis-Pons-RomanRoy1986}.
Indeed, applying $\frac{\partial}{\partial\dot q}$ to
\bref{pullback} we get
$$
W_{ij}\left(\frac{\partial \phi_\mu}{\partial
p_j}\right)_{\!|_{p=\hat p}} =0\,,\ \ \forall\, q,\dot q\,.
$$

The basis of null vectors $ \gamma_\mu$, with components $\gamma
_\mu ^j$, is denoted as
\beq\gamma_\mu^j := {\mathcal F}\!L^*\frac{\partial \phi_\mu}{\partial
p_j}\,.\label{gamma}
\eeq

Working with this basis proves to be an efficient way to obtain
results for the Lagrangian tangent space formulation by use of
Hamiltonian techniques.

\section{Expanding the Lagrangian around a classical solution \label{expanding}}

Denote the solution by $ q^o$. We have assumed that $q^o$ is
non-degenerate, i.e., the equations of motion $[L]$ have simple
zeroes in $q=q^o$. Although this condition is fulfilled for the
most of solutions in various models, there are Lagrangians with
degenerate solutions leading, for example, to ineffective
(irregular) constraints
\cite{Garcia-Pons1998,Miskovic-Zanelli,Miskovic-Troncoso-Zanelli}.

In the Hamiltonian formalism, the criterion to have all
constraints effective or functionally independent in the vicinity
of the solution $q^o$, is that their Jacobian in the phase space
$(q,p)$ evaluated at $(q^o,p^o)$, has maximal rank \cite{Dirac}.
This condition can be generalized to Lagrangian formalism. The $n$
Euler-Lagrange equations $[L]$, containing the evolution equation
(with non-vanishing Hessian) generally imply the existence of
primary Lagangian constraints (we introduce them
below). Typically, its preservation will yield new, secondary
constraints, then tertiary, and so on. In order to have full control
of the quadratic fluctuations theory around a solution $q^o$ we
will require that (\emph{i}) the rank of the Hessian matrix be
constant, (\emph{ii}) the equations of motion $[L]$ to have simple
zeroes in $q=q^o$ (the solutions are non-degenerate) and
(\emph{iii}) that all constraints are effective in a neighborhood
of $q^o$ (that is, its Jacobian with respect to the tangent space
coordinates be of maximum rank). Note that some of these
requirements may cease to hold only in one singular point, which
can therefore pass unnoticed if the given conditions are not
explicitly checked at this point. This happens in Chern-Simons
gauge theories, which have been discussed in Hamiltonian formalism
in
\cite{Saavedra-Troncoso-Zanelli,Miskovic-Zanelli,Miskovic-Troncoso-Zanelli}.

\vspace{6mm}

Now we introduce the fluctuations theory Lagrangian.
First define the fluctuations $Q$ by means of \beq q= q^o +
\epsilon Q \,,\qquad (\Rightarrow \dot q= \dot q^o + \epsilon \dot
Q )\,, \label{q-expansion} \eeq with $\epsilon$ a small constant
parameter, and expand
\begin{eqnarray}
L(q,\dot q) &=& L( q^o,\dot  q^o) + \epsilon ( Q\,\frac{\partial L}{\partial q}|_o
+ \dot Q\,\frac{\partial L}{\partial \dot q}|_o) +
\epsilon^2 \tilde L(Q,\dot Q;t) +
{\cal O}(\epsilon^3)
\nonumber \\
&=& L( q^o,\dot  q^o) + \epsilon
\frac{d}{dt}\,(Q \,\frac{\partial L}{\partial \dot q}|_o) + \epsilon^2 \tilde L(Q,\dot Q;t)
+{\cal O}(\epsilon^3)\label{ele}\,,
\end{eqnarray}
where\footnote{Note that in field theory the second term in
\bref{ele} is a divergence.} we have generically denoted $A(q,\dot
q)|_o = A(q^o,\dot q^o)$, recalled that $[L]|_o =0$ (the omitted
indices are saturated in an obvious way), and defined the
quadratic Lagrangian for small fluctuations\footnote{The size of
the fluctuations depends not only on the values of the $Q$
variables, but also on the \textquotedblleft
small\textquotedblright\ parameter $\epsilon$, which has been
factored out from the fluctuations theory.} \beq \tilde L(Q,\dot
Q; t):= \frac{1}{2} \Big( Q\frac{\partial^2L}{\partial q\partial
q}|_o Q + 2Q\frac{\partial^2L}{\partial q\partial \dot q}|_o \dot
Q + \dot Q\frac{\partial^2L}{\partial \dot q\partial \dot q}|_o
\dot Q \Big)\,. \label{tildeL} \eeq (Since this Lagrangian leads
to linear equations of motion, it is just the Lagrangian for
linearized fluctuations.) In general $\tilde L$ will be time
dependent because the solution $q^o(t)$ depends on time
explicitly. However, in order not to burden the notation, from now
on this time dependence will not be made explicit in the arguments
of our functions. Note that the Hessian matrix for $\tilde L$
coincides with the Hessian matrix for $L$ computed on the solution
$q^o$, $W|_o$. If we now perform a change of variables $q
\rightarrow Q$ (see the discussion on the change of variables in
Appendix \ref{expansions}), noticing that
$\frac{\partial}{\partial q}=
\frac{1}{\epsilon}\frac{\partial}{\partial Q}$ and
$\frac{\partial}{\partial \dot q}=
\frac{1}{\epsilon}\frac{\partial}{\partial \dot Q}$ and applying
\bref{ele}, for the Euler-Lagrange equations we obtain
\begin{eqnarray}
[L(q, \dot q)]_q &=& \frac{1}{\epsilon}\,[L(q^o + \epsilon Q \,,
\dot q^o + \epsilon \dot Q)]_Q = \frac{1}{\epsilon}\,(\epsilon^2
[\tilde L(Q,\dot Q)]_Q +{\cal O}(\epsilon^3))\nonumber \\
&=& \epsilon \,[\tilde L(Q,\dot Q)]_Q +{\cal O}(\epsilon^2)\,.
\label{eom-relat}
\end{eqnarray}
Thus we see that {\it  if $Q(t)$ is a
solution of the Euler-Lagrange e.o.m. for
$\tilde L$, then $q(t):= q^o(t) + \epsilon Q(t)$ is a solution of
the e.o.m. for the original Lagrangian $L$ up to terms of order
$\epsilon^2$}.

It can be verified that all quadratic first order Lagrangians
(which in general contain linear terms) are equivalent to their
own fluctuations theories. This can be seen by making the
coordinate transformation $q=q^{o}+q^{\prime }$, where $q^{o}$ is
a particular solution of the e.o.m., which transforms the
quadratic Lagrangian $L(q,\dot{q})$ into a homogenous function $L(
q^{o},\dot{q}^{o})+L_{hom}(q^{\prime },\dot{q}^{\prime })$ (up to
a total derivative), where $L_{hom}$ stands for the (second order)
homogenous part of $L$. Using the property of homogenous functions
of second degree, $L_{hom}(\epsilon Q)=\epsilon ^{2}L_{hom}(Q)$,
we conclude that the fluctuations Lagrangian is the homogenous
part of the original Lagrangian, $
\tilde{L}(Q,\dot{Q})=L_{hom}(Q,\dot{Q})\,. $ This is valid exactly
(for any $\epsilon $). The fluctuations Lagrangian and the
original quadratic Lagrangian have, therefore, equivalent
dynamical structures, including both gauge and rigid symmetries.

\subsection{Lagrangian constraints}

The equations $[L]$ can be separated into the evolution equations, and the
constraints.
Taking into account that $W\gamma_\mu =0$ identically,
the primary Lagrangian constraints for $L$ are
\beq
\chi_\mu := [L]\gamma_\mu =
(\alpha -W\ddot q) \gamma_\mu =  \alpha  \gamma_\mu \simeq 0\,,
\label{prim-lagr-constr}
\eeq
where $\simeq 0$ means \textquotedblleft vanishing on shell\textquotedblright ,
that is, when the e.o.m. are satisfied.
If we expand them under $q= q^o + \epsilon Q$ we get (note that
$\chi_\mu(q^o, \dot q^o)=0$ because of $[L]|_o=0$),
\begin{eqnarray}
\chi_\mu &=&(\alpha  \gamma_\mu)(q, \dot q)=
\epsilon (Q\,\frac{\partial (\alpha  \gamma_\mu)}{\partial q}|_o
+ \dot Q \,\frac{\partial (\alpha  \gamma_\mu)}{\partial \dot q}|_o)
+{\cal O}(\epsilon^2) \nonumber \\
&=& \epsilon (Q\, \frac{\partial ([L] \gamma_\mu)}{\partial q}|_o
+ \dot Q\, \frac{\partial ([L] \gamma_\mu)}{\partial \dot q}|_o)
+{\cal O}(\epsilon^2)\nonumber \\
&=& \epsilon (Q\, (\frac{\partial [L] }{\partial q}\,\gamma_\mu)|_o
+ \dot Q\, (\frac{\partial [L] }{\partial \dot q}\,\gamma_\mu)|_o)
+{\cal O}(\epsilon^2) =: \epsilon\tilde \chi_\mu +{\cal O}(\epsilon^2)\,.
 \end{eqnarray}

This result suggests that
\beq
\tilde \chi_\mu := Q \,(\frac{\partial [L] }{\partial q}\,\gamma_\mu)|_o
+ \dot Q\, (\frac{\partial [L] }{\partial \dot q}\,\gamma_\mu)|_o\simeq 0
\label{lagrConstrQ}\eeq
are the primary Lagrangian constraints
for the theory derived from $\tilde L$.
Now we will prove this claim.

To this end, notice that we can expand directly $[L(q, \dot q)]$,
$$[L(q, \dot q)]_q =[L(q, \dot q)]_{q^o} +
\epsilon \,( Q \,\frac{\partial[L]}{\partial q}|_o
+ \dot Q \,\frac{\partial[L]}{\partial \dot q}|_o +
\ddot Q \,\frac{\partial[L]}{\partial \ddot q}|_o)+{\cal O}(\epsilon^2)\,,
$$
and use that $[L(q, \dot q)]_{q^o}=0$ by definition. Then, comparing
the above expression with \bref{eom-relat},
we conclude that the Euler-Lagrange e.o.m.  for $\tilde L$ can be written as
$$
[\tilde L(Q,\dot Q)] = Q\, \frac{\partial[L]}{\partial q}|_o
+ \dot Q\, \frac{\partial[L]}{\partial \dot q}|_o - W\!|_o \ddot Q\,.
$$
Thus, using the fact that $\gamma_\mu\!|_o$ are the null vectors of
the Hessian $W\!|_o$ for $\tilde L$, the primary Lagrangian constraints
for $\tilde L$ are
$$
\Big([\tilde L(Q,\dot Q)]\gamma_\mu\Big)\!|_o = (Q \,\frac{\partial[L]}{\partial q}|_o
+ \dot Q\, \frac{\partial[L]}{\partial \dot q}|_o)\gamma_\mu|_o= \tilde \chi_\mu\,,
$$
which coincides with the result in \bref{lagrConstrQ} and proves the claim.

\vspace{6mm}

The number of the constraints $\tilde{\chi}_{\mu }$ ($\mu
=1,\ldots ,k$) is the same as the number of the original constraints $\chi
_{\mu }$ since we are dealing with effective constraints, for which the
Jacobian $\frac{\partial (\chi _{1},\ldots ,\chi _{k})}{\partial (q,\dot{q})}|_o
$ has be non degenerate (has rank $k$). From this, it follows immediately
that $\tilde{\chi}_{1},\ldots ,\tilde{\chi}_{k}$ are linearly independent.

\section{The canonical formalism}

This result $\chi_\mu=\epsilon \tilde \chi_\mu+{\cal
O}(\epsilon^2)$ makes one suspect that the full algorithm of
constraints for the original theory will be reproduced, step by
step, within the theory of linearized fluctuations. On the other
hand we know that the Lagrangian and Hamiltonian constraint
algorithms are deeply related, see
\cite{Batlle-Gomis-Pons-RomanRoy1986,Pons:1986zg}, in the sense
that, step by step, one can determine a subset of the Lagrangian
constraints as pullbacks ---under the Legendre map--- of the
Hamiltonian constraints, and the rest from the canonical
determination of some of the arbitrary functions that appear as
Lagrange multipliers in the Dirac Hamiltonian. Since the analysis
of the constraint algorithm in the canonical formalism is
facilitated by the presence of the Poisson bracket structure, we
now turn to the canonical analysis.

If we use the change of variables $q\rightarrow Q$, then $\hat p$
becomes $\hat p =\frac{1}{\epsilon}\frac{\partial L}{\partial \dot
Q}$ and, using the expansion \bref{ele}, we obtain
\beq
\hat p = \frac{1}{\epsilon}\Big(\epsilon \frac{\partial L}{\partial \dot
q}|_o + \epsilon^2 \frac{\partial \tilde L}{\partial \dot Q} +
{\cal O}(\epsilon^3)\Big)=: p^o + \epsilon\hat P +
\epsilon^2 F+{\cal O}(\epsilon^3)\,,
\label{hat-p-expansion}
\eeq
where $p^o:= \hat
p(q^o,\dot q^o)$ are the momenta corresponding to the solution of
the e.o.m. and $F(Q, \dot Q)$ are functions quadratic in $Q, \dot
Q$ that can be easily determined. Note that $\hat P$ define the
Legendre map for the theory of linearized fluctuations. We see
that at first order in $\epsilon$ the expansion for $\hat p$
behaves as expected.

The canonical Hamiltonian $H(q,p)$ associated with the Lagrangian
$L$ is characterized by  $H(q,\hat p)=\hat p \dot q - L$. It was
shown by Dirac that this function always exists and it is only
determined up to the addition of primary Hamiltonian constraints
$\phi_\mu$.

The fluctuation momenta $P$ are defined in the canonical formalism
from
\beq
p =: p^o+\epsilon P\,.
\label{p-expansion}
\eeq
Comparing this expansion with the expansion \bref{hat-p-expansion},
we find that the pullback map $p\rightarrow \hat{p}$ implies, under the
change of variables \bref{q-expansion} and \bref{p-expansion},
the map $P\rightarrow \hat{P}+\epsilon F(Q,\dot{Q})+\mathcal{O}(\epsilon ^{2})$, that is,
\beq
p\rightarrow \hat{p}\ \ \Rightarrow \
P\rightarrow \hat{P}+\epsilon F(Q,\dot{Q})+\mathcal{O}(\epsilon ^{2})\,.
\label{p-pullbackbis}
\eeq
which is different from the pullback map in the canonical fluctuations theory
$P\rightarrow \hat{P}$. This mismatch between the two pullback operations
---for the original theory and for the fluctuations theory--- is of order $\epsilon$
and has no consequences as regards the mutual consistency of the Lagrangian and
Hamiltonian version of the fluctuations theory.

Now consider the expansion for the primary Hamiltonian
constraints,
\begin{eqnarray}
\phi_\mu(q,p) &=& \phi_\mu(q^o,p^o)
+\epsilon(Q \frac{\partial \phi_\mu}{\partial q}|_o
+ P \frac{\partial \phi_\mu}{\partial p}|_o) + \epsilon^2 B_\mu(Q,P)
+  {\cal O}(\epsilon^3)\nonumber \\
&=:& \epsilon\tilde\phi_\mu(Q,P)+ \epsilon^2 B_\mu(Q,P)+ {\cal
O}(\epsilon^3)\,, \label{phi-expansion}\end{eqnarray} which,
again, suggests that \beq\tilde\phi_\mu := Q \frac{\partial
\phi_\mu}{\partial q}|_o + P \frac{\partial \phi_\mu}{\partial
p}|_o \label{tildephi} \eeq are the primary constraints for the
canonical theory of linearized fluctuations. $B_\mu(Q,P)$ are
functions quadratic in $Q,P$,
 \beq
B_\mu(Q,P):= \frac{1}{2}
\Big( Q\frac{\partial^2\phi_\mu}{\partial q\partial q}|_o Q +
2Q\frac{\partial^2\phi_\mu}{\partial q\partial p}|_o P +
P\frac{\partial^2\phi_\mu}{\partial p\partial p}|_o P \Big)\,.
\label{2phimu}
\eeq
\subsection{Primary constraints}

Let us verify that $\tilde\phi_\mu$ are indeed the primary
Hamiltonian constraints for the theory originating in the
fluctuations Lagrangian $\tilde L$. Since $\phi_\mu(q,\hat p)=0$
identically\footnote{Note that $\phi _\mu (q,p) \simeq 0$
(the constraints vanish on the constraint surface, but their
derivatives not), while $\phi _\mu (q,\hat p) = 0$
identically.},
we also have $$\frac{\partial \phi_\mu}{\partial
q}|_o + \frac{\partial \phi_\mu}{\partial p}|_o \frac{\partial
\hat p}{\partial q}|_o =0\,, $$ which implies
$$
\tilde\phi_\mu(Q,P) = (P-Q\frac{\partial \hat p}{\partial q}|_o)
\frac{\partial \phi_\mu}{\partial p}|_o =
\Big((P-Q\frac{\partial L}{\partial q\partial \dot q})\gamma_\mu\Big)|_o\,.
$$
Now one can check that $\tilde\phi_\mu(Q,\hat P)=0$ identically. Indeed,
$
\hat P (Q, \dot Q) = \frac{\partial \tilde L}{\partial \dot Q}\,,
$
and using \bref{tildeL},
$$
\hat P (Q, \dot Q)= Q \frac{\partial^2  L}{\partial q\partial \dot q}|_o+
\dot Q W\!|_o\,.
$$
Now, since $\gamma_\mu|_o$ are the null vectors of the Hessian matrix $W\!|_o$,
we obtain that
$$\Big((\hat P-Q\frac{\partial L}{\partial q\partial \dot q})\gamma_\mu\Big)|_o =0$$
identically, which proves that indeed $\tilde\phi_\mu$ are the
primary constraints for the canonical theory of linearized
fluctuations.

\subsection{The canonical Hamiltonian}
\label{subsec:H}

Here we find the quadratic Hamiltonian for the linearized fluctuations.
Consider the canonical Hamiltonian $H(q,p)$\footnote{As said before, the
canonical Hamiltonian is not unique, but any choice satisfying
$H(q,\hat p)= \hat p \dot q - L(q,\dot q)$ will work.} and expand it
in $\epsilon$. This will define a candidate $\bar H(Q,P)$
for the quadratic canonical Hamiltonian of linearized fluctuations, but
it should be checked that $\bar H(Q,\hat P) = \hat P\dot Q - \tilde L$.
We will see that it is not exactly so. The reason is that, as Dirac
already emphasized, the true Hamiltonian dynamics is
described by the Dirac Hamiltonian
\beq
H_D(q,p):= H(q,p)+\lambda^\mu\phi_\mu\,.
\label{HDirac}
\eeq
In order to find the quadratic canonical Hamiltonian for fluctuations theory,
let us first expand the canonical Hamiltonian,
$$
H(q,p)=: H(q^o ,p^o) + \epsilon (Q\frac{\partial H}{\partial q}|_o
+ P\frac{\partial H}{\partial p}|_o ) + \epsilon^2\bar H(Q,P)+
{\cal O}(\epsilon^3)\,,
$$
where $\bar H(Q,P)$ is quadratic in $Q,P$, \beq
 \bar H(Q,P):= \frac{1}{2}
\Big( Q\frac{\partial^2H}{\partial q\partial q}|_o Q +
2Q\frac{\partial^2H}{\partial q\partial p}|_o P +
P\frac{\partial^2H}{\partial p\partial p}|_o P \Big)\,.
\label{barH}
\eeq

Since $q^o, p^o$ satisfy the e.o.m.,
\begin{eqnarray}
\dot q^o&=& \frac{\partial H}{\partial p}|_o +
\lambda^\mu(q^o,\dot q^o)\frac{\partial \phi_\mu}{\partial p}|_o \nonumber \\
\dot p^o&=& -\frac{\partial H}{\partial q}|_o
-\lambda^\mu(q^o,\dot q^o)\frac{\partial \phi_\mu}{\partial q}|_o\label{eomH}\,,
\end{eqnarray}
where the Lagrange multipliers $\lambda^\mu$ can always be
determined as definite functions in tangent space by using the e.o.m.
for $q$ and the pullback $p\rightarrow \hat p$ (see Appendix \ref{multipliers}
for more details), we can replace
\begin{eqnarray}
H(q,p) &=& H(q^o ,p^o) +\epsilon \Big(Q(-\dot
p^o-\lambda^\mu(q^o,\dot q^o)\frac{\partial \phi_\mu}{\partial
q}|_o) + P(\dot q^o-\lambda^\mu(q^o,\dot q^o)\frac{\partial
\phi_\mu}{\partial p}|_o) \Big)
\nonumber \\
&+& \epsilon^2\bar H(Q,P)+ {\cal O}(\epsilon^3)\nonumber \\
&=& H(q^o ,p^o) +\epsilon (P\dot q^o -Q\dot p^o) -\epsilon
\lambda^\mu(q^o,\dot q^o)\tilde\phi_\mu + \epsilon^2\bar H(Q,P)+
{\cal O}(\epsilon^3)\,,\label{hcan-exp}
\end{eqnarray}
where the definition \bref{tildephi} has been used. Now consider
the expansion for the functions $\lambda^\mu(q,\dot q)$,
$$
\lambda^\mu(q,\dot q) = \lambda^\mu(q^o,\dot q^o)
+ \epsilon\tilde \lambda^\mu(Q,\dot Q)+ {\cal O}(\epsilon^2)\,,$$
and recall \bref{phi-expansion}. Then,
the Dirac Hamiltonian \bref{HDirac} has the expansion
\begin{eqnarray}
H_D(q,p)&=&H(q^o ,p^o)  +\epsilon (P\dot q^o -Q\dot p^o)
 \nonumber \\
&+& \epsilon^2 \Big( \bar H(Q,P) + \lambda^\mu(q^o,\dot
q^o)B_\mu(Q, P) + \tilde \lambda^\mu \tilde\phi_\mu(Q, P) \Big) +
{\cal O}(\epsilon^3)\,. \label{hd}
\end{eqnarray}
Unlike in the expansion of $L$ given by eq. \bref{ele},
where the linear term does not contribute to the Lagrangian e.o.m., now
the linear term in $H_{D}$ does contribute to the Hamiltonian e.o.m.
At the end of this subsection we will see that this is
consistent with the Hamilton's equations of the original theory.

Expecting that the quadratic Dirac Hamiltonian of linearized fluctuations
has the usual form $\tilde H_D (Q,P) =\tilde H(Q,P)+
\tilde\lambda^\mu (Q,\dot Q) \tilde\phi_\mu (Q,P)$,
the result \bref{hd} strongly suggests that the true
canonical Hamiltonian for the fluctuations
is  not $\bar H(Q,P)$, but the whole expression
$\bar H(Q,P) +  \lambda^\mu(q^o,\dot q^o)B_\mu(Q, P)$.
We shall prove this assertion
in the following. Equation \bref{p-pullbackbis} allows
us to substitute $\hat p$ for $p$
in the l.h.s. of \bref{hd}, and $\hat P + \epsilon F + {\cal O}(\epsilon^2)$ for $P$
in the r.h.s. Taking into account that $H_D(q,\hat p)= H(q,\hat p)$
for $\phi (q,\hat p)=0$, we obtain
\begin{eqnarray}
H_D(q,\hat p)&=&  H(q^o ,p^o)  +\epsilon (\hat P\dot q^o -Q\dot
p^o)
\nonumber \\
&+& \epsilon^2 \Big(\bar H(Q,\hat P) + \lambda^\mu(q^o,\dot
q^o)B_\mu(Q, \hat P) + \dot q^o F(Q,\dot Q)\Big) + {\cal
O}(\epsilon^3)\,. \label{compare1}
\end{eqnarray}
This expression must be compared with what we obtain directly from
the fact that $H(q,\hat p)=\hat p \dot q - L$. Using
\bref{p-expansion} and \bref{ele}
\begin{eqnarray}
H(q,\hat p)&=& \hat p \dot q - L(q,\dot q) =
\Big(p^o + \epsilon\hat P + \epsilon^2 F+ {\cal O}(\epsilon^3)\Big)
( \dot q^o + \epsilon\dot Q)\nonumber \\
&-& \Big(L( q^o,\dot  q^o) + \epsilon \frac{d}{dt}(Q p^o) +
\epsilon^2 \tilde L(Q,\dot Q )
+{\cal O}(\epsilon^3)\Big)\nonumber \\
&=& p^o\dot q^o - L( q^o,\dot  q^o) + \epsilon (\hat P \dot q^o - Q \dot p^o )
+ \epsilon^2 ( \hat P \dot Q - \tilde L + \dot q^o F )+ {\cal O}(\epsilon^3)
\nonumber \\
&=:& H(q^o ,p^o)  +\epsilon (\hat P\dot q^o -Q\dot p^o)+\epsilon^2(\tilde H(Q,\hat P)
+ \dot q^o F )+ {\cal O}(\epsilon^3)\,,\label{compare2}
\end{eqnarray}
where we have defined the true canonical Hamiltonian $\tilde H(Q, P)$ such that
$\tilde H(Q,\hat P)=\hat P \dot Q - \tilde L$.

Now we compare \bref{compare1} and \bref{compare2}. It follows that
$$\tilde H(Q,\hat  P)= \bar H(Q,\hat  P)+
\lambda^\mu(q^o,\dot q^o)B_\mu(Q,\hat P)\,,
$$
and hence the canonical quadratic Hamiltonian for the fluctuations is
\beq
\tilde H(Q, P)= \bar H(Q, P)+
\lambda^\mu(q^o,\dot q^o)B_\mu(Q,P)\,.
\label{H-tilde}
\eeq
This proves our assertion.

Now \bref{compare1}
can be written as
\beq
H_D =  H(q^o ,p^o)  +\epsilon ( P\dot q^o -Q\dot p^o)
+ \epsilon^2
\tilde H_D
 + {\cal O}(\epsilon^3)\,,
\label{hdhd}
\eeq
with $\tilde H_D = \tilde H + \tilde  \lambda^\mu\tilde \phi_\mu\,.$

Now we can state the following result: {\sl if $Q(t), P(t)$ is a solution
of the Hamilton-Dirac's equations for
the fluctuation dynamics, then $q(t) := q^o(t) + \epsilon Q(t)\,,\ \
p(t) := p^o(t) +\epsilon P(t)$ is a solution
of the Hamilton's equations
for the original dynamics up to terms of order $\epsilon^2 $.}

To prove it just consider the equations of the original dynamics and use
\bref{hdhd},
\begin{eqnarray}
\dot q &=&
\frac{1}{\epsilon}\frac{\partial H_D}{\partial P}= \frac{1}{\epsilon}
\Big(\epsilon \dot q^o +
\epsilon^2\frac{\partial\tilde  H_D}{\partial P} + {\cal O}(\epsilon^3)\Big)
=  \dot q^o +
\epsilon\frac{\partial\tilde  H_D}{\partial P} + {\cal O}(\epsilon^2)\,,
\nonumber\\
\dot p &=&
-\frac{1}{\epsilon}\frac{\partial H_D}{\partial Q}= -\frac{1}{\epsilon}
\Big(-\epsilon \dot p^o +
\epsilon^2\frac{\partial\tilde  H_D}{\partial Q} + {\cal O}(\epsilon^3)\Big)
 = \dot p^o -
\epsilon\frac{\partial\tilde  H_D}{\partial Q} + {\cal O}(\epsilon^2)\,.
\nonumber
\end{eqnarray}

Since the equations for the fluctuation dynamics are
\beq\dot Q = \frac{\partial\tilde  H_D}{\partial P}\,,\qquad\qquad
\dot P = -\frac{\partial\tilde  H_D}{\partial Q}\,,
\eeq
the result follows.
\subsection{The algebra of constraints}

The change of variables $q\rightarrow Q,\ p\rightarrow P$ is canonical
up to a factor  $\epsilon^2$. We define a new bracket for the theory of
fluctuations by
$$\{-,\,-\}^{\tilde{}}= {\epsilon^2}\{-,\,-\}$$
in order to have
$\{Q,\,P\}^{\tilde{}} = \{q,\,p\}= \delta$.

Let us also define the auxiliary differential operator $D^h$, acting on
functions of the original variables $q,p$, as
$D^h:= (Q\frac{\partial }{\partial q}+P\frac{\partial }{\partial p})|_o$,
so that, for any $f(q, p; t)$, it gives the first order term in the expansion,
$f=f|_o + \epsilon\, D^hf + {\cal O}(\epsilon^2)$,
where $f|_o =f(q^o, p^o; t)$.
Now consider the constraints $\tilde \phi_\mu$. Let us first
evaluate their Poisson brackets. Using \bref{ap1} we find
\bea
\{\tilde \phi_\mu,\,\tilde \phi_\nu\}^{\tilde{}}=
\{ \phi_\mu,\,\phi_\nu\}|_o\,.\label{equalbrackets}
\eea
Thus the structure ---First Class, Second Class---
of the primary constraints is fully inherited in the
fluctuations formalism and is fixed \textquotedblleft on shell\textquotedblright\ .
Suppose that the original primary constraints $\phi_\mu\equiv\phi_\mu^{(0)} $
(the superindex ${}^{(0)}$ is for primary) split into First Class
constraints $\phi_{\mu_0}^{(0)}$ and Second Class constraints $\phi_{\mu'_0}^{(0)}$.
Then the secondary constraints are obtained as
$\phi_{\mu_0}^{(1)}:= \{\phi_{\mu_0}^{(0)},\, H\}$. Now we proceed in the same way with
the fluctuations theory.
The same splitting repeats for the constraints $\tilde \phi_\mu^{(0)}$ .
The only difference in finding the secondary constraints is that
$\tilde \phi_\mu^{(0)}$
are in general time dependent.

Then, using the form of the Hamiltonian \bref{H-tilde} and
the fact that it is always $\tilde\phi=D^h\phi$ (for any indices), for the
secondary constraints of the fluctuations theory we obtain
\begin{eqnarray}
\tilde\phi_{\mu_0}^{(1)}&:=&\frac{\partial}{\partial t}\tilde \phi_{\mu_0}^{(0)}+
\{\tilde\phi_{\mu_0}^{(0)},\, \tilde H \}^{\tilde{}}
=Q \frac{d}{d t}\frac{\partial \phi_{\mu_0}^{(0)}}{\partial q}|_o
+ P \frac{d}{d t}\frac{\partial \phi_{\mu_0}^{(0)}}{\partial p}|_o
+\{\tilde\phi_{\mu_0}^{(0)},\, \bar H \}^{\tilde{}} \nonumber \\ &+&
\lambda^\mu(q^o,\dot q^o)
\{\tilde\phi_{\mu_0}^{(0)},\, B_\mu \}^{\tilde{}}\,.
\end{eqnarray}
The bracket in the last term can be transformed with the help of the
expansion \bref{phi-expansion} and the identity \bref{ap2}
(see Appendix \ref{expansions})
\beq\{\tilde\phi_{\mu_0}^{(0)},\, B_\mu \}^{\tilde{}} =
\{\tilde\phi_\mu^{(0)},\, B_{\mu_0} \}^{\tilde{}} +
D^h\{\phi_{\mu_0}^{(0)},\, \phi_{\mu}^{(0)} \}\,.\label{solved}
\eeq
Since $\phi_{\mu_0}^{(0)}$ are First Class constraints,
$\{\phi_{\mu_0}^{(0)},\, \phi_{\mu}^{(0)} \} =
\alpha^\nu_{\mu_0\mu}\phi_{\nu}^{(0)}$ for some functions
$\alpha^\nu_{\mu_0\mu}$. Therefore, \beq
D^h\{\phi_{\mu_0}^{(0)},\, \phi_{\mu}^{(0)} \} =
D^h(\alpha^\nu_{\mu_0\mu}\phi_{\nu}^{(0)})=\alpha^\nu_{\mu_0\mu}|_oD^h\phi_{\nu}^{(0)}
= \alpha^\nu_{\mu_0\mu}|_o\tilde\phi_{\nu}^{(0)}\simeq 0\,, \eeq which
shows that the last term in \bref{solved} vanishes on the surface
of primary constraints (this is the meaning of $\simeq$ at this
stage).

Using the previous results and the e.o.m. \bref{eomH},
a little computation shows that
\begin{eqnarray}
&&
Q \frac{d}{d t}\frac{\partial \phi_{\mu_0}^{(0)}}{\partial q}|_o
+ P \frac{d}{d t}\frac{\partial \phi_{\mu_0}^{(0)}}{\partial p}|_o
+ \lambda^\mu(q^o,\dot q^o)
\{\tilde\phi_{\mu_0}^{(0)},\, B_\mu \}^{\tilde{}}\nonumber
\\
&\simeq&
Q \frac{d}{d t}\frac{\partial \phi_{\mu_0}^{(0)}}{\partial q}|_o
+ P \frac{d}{d t}\frac{\partial \phi_{\mu_0}^{(0)}}{\partial p}|_o
+ \lambda^\mu(q^o,\dot q^o)
\{\tilde\phi_\mu^{(0)},\, B_{\mu_0} \}^{\tilde{}}\nonumber \\ &=&
Q\{\frac{\partial \phi_{\mu_0}^{(0)}}{\partial q},\,H\}|_o +
P\{\frac{\partial \phi_{\mu_0}^{(0)}}{\partial p},\,H\}|_o\,,\label{little1}
\end{eqnarray}
whereas
\beq\{\tilde\phi_{\mu_0}^{(0)},\, \bar H \}^{\tilde{}}=
Q\{\phi_{\mu_0}^{(0)},\,\frac{\partial H}{\partial q}\}|_o +
P\{\phi_{\mu_0}^{(0)},\,\frac{\partial H}{\partial p}\}|_o \,.\label{little2}
\eeq
Altogether gives
$$\tilde\phi_\mu^{(1)}\simeq
(Q\frac{\partial }{\partial q}+P\frac{\partial }{\partial p})
\{\phi_{\mu_0}^{(0)},\, H\}|_o
=(Q\frac{\partial }{\partial q}+P\frac{\partial }{\partial p})
\phi_{\mu_0}^{(1)}|_o = D^h\phi_{\mu_0}^{(1)}\,,
$$

Note the complete analogy of the expansion of secondary constraints
with that of the primary constraints
\bref{phi-expansion}. Namely, there we had the expansion
 $$\phi_\mu^{(0)}(q,p) =
\epsilon\tilde\phi_\mu^{(0)}(Q,P)+{\cal O}(\epsilon^2),$$ now we find
\beq\phi_{\mu_0}^{(1)}(q,p) \simeq
\epsilon\tilde\phi_{\mu_0}^{(1)}(Q,P)+{\cal O}(\epsilon^2)\,.
\label{phi-expansion2}\eeq

On the other hand, recalling Appendix \ref{expansions}, \beq
\{D^hf,\,D^hg\}^{\tilde{}}=\{f,\,g\}|_o\,. \label{equalbrackets2}
\eeq
Thus {\sl our algebra of primary and
secondary constraints for the fluctuations theory just mimics the
algebra of the original constraints computed at $q^o, p^o$}. This
means in particular that the First Class constraints become
Abelianized for the fluctuations theory.

In Sec.\ref{expanding} we showed that the original and
fluctuations Lagrangians have equivalent dynamical structures
in the case of quadratic Lagrangians.
Thus,
the fact that the fluctuations Lagrangian cannot have non-Abelian
symmetry means that non-Abelian symmetry cannot be described
by a quadratic Lagrangian.

\subsection{Dirac brackets}

The constraint algorithm now continues in parallel for the
original theory and for the fluctuations theory. Let us relate the Dirac brackets
at the level of the primary constraints for both theories. The matrix of
Second Class constraints
$$\{\tilde\phi_{\mu'_0}^{(0)},\,\tilde\phi_{\nu'_0}^{(0)}\}^{\tilde{}}=
\{D^h\phi_{\mu'_0}^{(0)},\,D^h\phi_{\nu'_0}^{(0)}\}^{\tilde{}}=
\{\phi_{\mu'_0}^{(0)},\,\phi_{\nu'_0}^{(0)}\}|_o=: M_{\mu'_0\nu'_0}|_o\,,
$$
is invertible. Consider the inverse $M^{\mu'_0\nu'_0}|_o$.
Dirac brackets for the fluctuations theory are then defined by
$$\{-\,,-\}^{\tilde{*}}:= \{-\,,-\}^{\tilde{}}
-\{-\,,\tilde\phi_{\mu'_0}^{(0)}\}^{\tilde{}}\,M^{\mu'_0\nu'_0}|_o
\,\{\tilde\phi_{\nu'_0}^{(0)}\,,-\}^{\tilde{}}
$$
Then, for any functions $f$, $g$ in the original phase space,
the following Dirac bracket can be calculated,
\begin{eqnarray}
\{D^hf\,,D^hg\}^{\tilde{*}}&:=& \{D^hf\,,D^hg\}^{\tilde{}}
-\{D^hf\,,\tilde\phi_{\mu'_0}^{(0)}\}^{\tilde{}}\,M^{\mu'_0\nu'_0}|_o \,
\{\tilde\phi_{\nu'_0}^{(0)}
\,,D^hg\}^{\tilde{}}\nonumber \\
&=&\{f\,,g\}|_o-\{f\,\phi_{\mu'_0}^{(0)}\}|_o\,M^{\mu'_0\nu'_0}|_o \,
\{\phi_{\nu'_0}^{(0)}\,,g\}|_o =\{f\,,g\}^*|_o\,.\label{dbrack}
\end{eqnarray}
Equation \bref{dbrack} is the analogous of \bref{equalbrackets2},
now for Dirac brackets.

The knowledge that the primary constraints $\tilde\phi_\mu^{(0)}$
and the secondary constraints $\tilde\phi_{\mu_0}^{(1)}$ for the
fluctuations theory are just $\tilde\phi_\mu = D^h\phi_\mu$ and
$\tilde\phi_{\mu_0}^{(1)}= D^h\phi_{\mu_0}^{(1)}$, together with
the results \bref{equalbrackets2} and \bref{dbrack}, allows to
continue the constraint algorithm another step. The same
parallelisms continue until the algorithm is finished. This proves
that {\sl the full algebra of constraints in the fluctuations
theory mimics the algebra of the original constraints computed at
$q^o, p^o$. In consequence,
the original theory and the fluctuations theory have the same
number of physical degrees of freedom.}
The Abelianization of the First Class constraints for
the fluctuations theory is a general phenomenon\footnote{In
fact gauge symmetries for quadratic systems are always Abelian
because the Hamiltonian constraints are linear --thus their
Poisson bracket is field independent-- and hence the only way to
exhibit First Class constraints is through the vanishing of their
Poisson brackets with all the constraints. Non-Abelian theories
and self-interaction --associated with terms in the action of
order higher than quadratic-- go hand in hand.}. Since
combinations of the First Class constraints generate gauge
symmetries, and since their number remains unchanged, the
dimensions of the original gauge group and the Abelian gauge group
in the fluctuations theory, are the same.

\subsection{Connection with the Lagrangian constraints}

Using results in \cite{Batlle-Gomis-Pons-RomanRoy1986},
the primary Lagragian constraints can be written as
\begin{eqnarray}\chi_{\mu_0}&=& {\mathcal F}\!L^*\{\phi_{\mu_0}^{(0)},\,H\}
= {\mathcal F}\!L^*\phi_{\mu_0}^{(1)}\label{chi}  \\
\chi_{\mu'_0}&=&{\mathcal F}\!L^*\{\phi_{\mu'_0}^{(0)},\,H\}+
\lambda^{\nu'_0}(q,\dot q)
{\mathcal F}\!L^*\{\phi_{\mu'_0}^{(0)},\, \phi_{\nu'_0}^{(0)}\}\label{chi'}\,.
 \end{eqnarray}
The rationale of \bref{chi} is that the pullback of a Hamiltonian
constraint must be a Lagrangian constraint. As for \bref{chi'} the
idea is that the time evolution of a ---now Second Class---
constraint must also vanish \textquotedblleft on shell\textquotedblright\ (here it is relevant that
the Lagrange multipliers $\lambda^{\nu'_0}$ are definite functions
in tangent space).

Let us use the notation\footnote{$D^l$ plays the same role in tangent space
as $D^h$ plays in phase space.}
$D^l:=(Q\frac{\partial }{\partial q}
+\dot Q\frac{\partial }{\partial \dot q})|_o$
and the fact, easily proved, that
$$D^l \circ {\mathcal F}\!L^* = \tilde{{\mathcal F}\!L^*}\circ D^h\,,$$
where $\tilde{{\mathcal F}\!L^*}$ is the pullback operation
$P\rightarrow \hat P$ for the fluctuations theory. Now expand
\bref{chi} in $\epsilon$, we get \beq\chi_{\mu_0}= \epsilon
D^l({\mathcal F}\!L^*\phi_{\mu_0}^{(1)})+ {\cal O}(\epsilon^2) =
\epsilon\tilde{{\mathcal F}\!L^*}(D^h\phi_{\mu_0}^{(1)})+ {\cal
O}(\epsilon^2)
 = \epsilon\tilde{{\mathcal F}\!L^*}\tilde\phi_{\mu_0}^{(1)}+ {\cal O}(\epsilon^2)\,,
 \eeq
which means that indeed the relation \bref{chi}
is preserved for the fluctuations theory as well, that is,
 $\tilde\chi_{\mu_0}=
\tilde{{\mathcal F}\!L^*}\tilde\phi_{\mu_0}^{(1)}$.

Let us do the same with \bref{chi'},
\begin{eqnarray}\chi_{\mu'_0}
&=&
\epsilon \Big( D^l({\mathcal F}\!L^*\{\phi_{\mu'_0}^{(0)},\,H\})
+ D^l(\lambda^{\nu'_0}
{\mathcal F}\!L^*\{\phi_{\mu'_0}^{(0)},\, \phi_{\nu'_0}^{(0)}\})\Big)
+ {\cal O}(\epsilon^2)\nonumber \\
&=&
\epsilon
\Big(\tilde{{\mathcal F}\!L^*}( D^h\{\phi_{\mu'_0}^{(0)},\,H\}) +
(D^l\lambda^{\nu'_0})
\{\phi_{\mu'_0}^{(0)},\,
\phi_{\nu'_0}^{(0)}\}|_o + \lambda^{\nu'_0}|_o
\tilde{{\mathcal F}\!L^*}( D^h\{\phi_{\mu'_0}^{(0)},\,
\phi_{\nu'_0}^{(0)}\})\Big)
+ {\cal O}(\epsilon^2)
\nonumber \\ &=&
\epsilon\Big(\tilde{{\mathcal F}\!L^*}
( D^h\{\phi_{\mu'_0}^{(0)},\,H+\lambda^{\nu'_0}|_o\phi_{\nu'_0}^{(0)}\}) + (D^l\lambda^{\nu'_0})
\{\phi_{\mu'_0}^{(0)},\,
\phi_{\nu'_0}^{(0)}\}|_o\Big)
+{\cal O}(\epsilon^2)\,.\label{chiprime}
\end{eqnarray}
Recalling that $D^l\lambda^{\nu'_0}=\tilde\lambda^{\nu'_0}$, the
form of the expansion \bref{chiprime} indicates that the objects
\beq
 \tilde{{\mathcal F}\!L^*}
(
D^h\{\phi_{\mu'_0}^{(0)},\,H+\lambda^{\nu'_0}|_o\phi_{\nu'_0}^{(0)}\})
+ \tilde\lambda^{\nu'_0} \{\phi_{\mu'_0}^{(0)},\,
\phi_{\nu'_0}^{(0)}\}|_o\,,\label{object}
\eeq
should be the
primary Lagrangian constraints $\tilde\chi_{\mu'_0}$ for the
fluctuations theory. Let us check that this is indeed the case.
Working only with the theory of fluctuations and using
\bref{solved} and the analogous of \bref{little1}, \bref{little2},
now applied to $\tilde \phi_{\mu'_0}^{(0)}$, we get
\begin{eqnarray}
\tilde\chi_{\mu'_0}&=&
\tilde{{\mathcal F}\!L^*}\Big(\frac{\partial}{\partial t}\tilde \phi_{\mu'_0}^{(0)}+
\{\tilde\phi_{\mu'_0}^{(0)},\, \tilde H \}^{\tilde{}}\Big)
+ \tilde\lambda^{\nu'_0}\tilde{{\mathcal F}\!L^*}
\{\tilde\phi_{\mu'_0}^{(0)},\,\tilde\phi_{\nu'_0}^{(0)}\}^{\tilde{}}\nonumber \\
&=&
\tilde{{\mathcal F}\!L^*}\Big(\frac{\partial}{\partial t}\tilde \phi_{\mu'_0}^{(0)}+
\lambda^\mu|_o \{\tilde\phi_{\mu'_0}^{(0)},\, B_\mu \}^{\tilde{}}\Big)
+ \tilde{{\mathcal F}\!L^*}\{\tilde\phi_{\mu'_0}^{(0)},\, \bar H \}^{\tilde{}}
 + \tilde\lambda^{\nu'_0}\tilde{{\mathcal F}\!L^*}
\{\tilde\phi_{\mu'_0}^{(0)},\,\tilde\phi_{\nu'_0}^{(0)}\}^{\tilde{}}
\nonumber \\ &=&
\tilde{{\mathcal F}\!L^*}\Big(\frac{\partial}{\partial t}\tilde \phi_{\mu'_0}^{(0)}+
\lambda^\mu|_o (\{\tilde\phi_\mu^{(0)},\, B_{\mu'_0} \}^{\tilde{}}
+ D^h\{\phi_{\mu'_0}^{(0)},\,\phi_\mu^{(0)}\})\Big)
+
\tilde{{\mathcal F}\!L^*}\{\tilde\phi_{\mu'_0}^{(0)},\, \bar H \}^{\tilde{}}
+\tilde\lambda^{\nu'_0}\tilde{{\mathcal F}\!L^*}
\{\tilde\phi_{\mu'_0}^{(0)},\,\tilde\phi_{\nu'_0}^{(0)}\}^{\tilde{}}
\nonumber \\ &=&
\tilde{{\mathcal F}\!L^*}
( D^h\{\phi_{\mu'_0}^{(0)},\,H\}) +\lambda^{\mu}|_o
\tilde{{\mathcal F}\!L^*}( D^h\{\phi_{\mu'_0}^{(0)},\,\phi_{\mu}^{(0)}\})
+ \tilde\lambda^{\nu'_0}
\{\phi_{\mu'_0}^{(0)},\,\tilde\phi_{\nu'_0}^{(0)}\}|_o
\nonumber \\ &=&
\tilde{{\mathcal F}\!L^*}
( D^h\{\phi_{\mu'_0}^{(0)},\,H\}) +\lambda^{\nu'_0}|_o
\tilde{{\mathcal F}\!L^*}( D^h\{\phi_{\mu'_0}^{(0)},\,\phi_{\nu'_0}^{(0)}\})
+ \tilde\lambda^{\nu'_0}
\{\phi_{\mu'_0}^{(0)},\,\tilde\phi_{\nu'_0}^{(0)}\}|_o\nonumber \\ &=&
\tilde{{\mathcal F}\!L^*}
( D^h\{\phi_{\mu'_0}^{(0)},\,H+\lambda^{\nu'_0}|_o\phi_{\nu'_0}^{(0)}\})
+ \tilde\lambda^{\nu'_0}
\{\phi_{\mu'_0}^{(0)},\,
\phi_{\nu'_0}^{(0)}\}|_o\,,
\end{eqnarray}
which is exactly \bref{object}.

The constraint algorithm in tangent space for the
fluctuations theory continues in the same way. The result is that for each
constraint $\chi$ of the original theory there is a constraint
$\tilde\chi$ in the fluctuations theory, that can be determined through the
expansion $\chi = \epsilon\tilde\chi +{\cal O}(\epsilon^2)$.
\section{Noether symmetries}

Noether symmetries of the action are those continuous
symmetries for which the infinitesimal transformation
$\delta q$ induced on
the Lagrangian $L$ gives a total derivative --a divergence in
field theory. They exhibit a conserved quantity --conserved
current in field theory-- $G$ such that the equality
\beq
[L]_q\delta q + \frac{d}{dt}G =0 \label{Noethersymmetry}
\eeq
holds identically. Let us $\epsilon$-expand \bref{Noethersymmetry} according
to \bref{q-expansion}, using the expansion \bref{eom-relat} which
includes the next order in $\epsilon$.
Note that
$\delta q= \delta q|_o + \epsilon D^l\delta q + {\cal
O}(\epsilon^2)$, and $G=G|_o + \epsilon D^lG +\epsilon^2 D^{2l}G +
{\cal O}(\epsilon^3)$, where we have introduced the notation
$D^{2l}f$ for the second order term in the expansion of any $f(
q,\dot q, t)$, \beq
 D^{2l}f = \frac{1}{2}
\Big( Q\frac{\partial^2f}{\partial q\partial q}|_o Q +
2Q\frac{\partial^2f}{\partial q\partial \dot q}|_o \dot Q +
\dot Q\frac{\partial^2f}{\partial \dot q\partial \dot q}|_o \dot Q   \Big)\,.
\label{D2l-f}
\eeq
Now the l.h.s. of \bref{Noethersymmetry} becomes
\begin{eqnarray}
[L]_q\delta q + \frac{d}{dt}G &=& \Big(\epsilon [\tilde L(Q,\dot
Q)]_Q +\epsilon^2[D^{2l}L]_Q+{\cal O}(\epsilon^3)\Big)\Big(\delta
q|_o + \epsilon D^l\delta q + {\cal O}(\epsilon^2)\Big)\nonumber
\\  &+& \frac{d}{dt}\Big(G|_o + \epsilon D^lG + \epsilon^2 D^{2l}G
+ {\cal O}(\epsilon^3)\Big)\nonumber  \\&=& \epsilon \Big([\tilde
L(Q,\dot Q)]_Q\delta q|_o+ \frac{d}{dt}D^lG\Big)
\nonumber\\ &+&
\epsilon^2 \Big( [\tilde L(Q,\dot Q)]_QD^l\delta q+
[D^{2l}L]_Q\delta q|_o + \frac{d}{dt}(D^{2l}G)\Big)\,,
\end{eqnarray}
where $\frac{d}{dt}G|_o$ vanishes because $G|_o$ is a conserved
quantity evaluated on a solution. Thus \bref{Noethersymmetry}
implies, to the lowest orders in the expansion, \beq[\tilde
L(Q,\dot Q)]_Q\delta q|_o+ \frac{d}{dt}D^lG=0 \label{noether1}
\eeq and \beq [\tilde L(Q,\dot Q)]_Q D^l\delta q+
[D^{2l}L]_Q\delta q|_o + \frac{d}{dt}(D^{2l}G)=0\,.
\label{noether2} \eeq

According to \bref{noether1}, the transformations $\delta Q$ defined by
\beq\delta Q := \delta q|_o
\label{finalNoether}
\eeq
produce a Noether symmetry for $\tilde L$, with a linear conserved
quantity
$D^lG$.
Note that this symmetry is trivial when the original symmetry preserves
the background, that is,
when $\delta q|_o=0$. However, in this case, the equation \bref{noether2}
takes the form of the conservation law \bref{Noethersymmetry}. Indeed,
when $\delta q|_o=0$, from eq. \bref{noether2}, the transformations
$\tilde\delta Q$ defined by
\beq
\tilde\delta Q := D^l\delta q
\label{finalNoether2}
\eeq
lead to a Noether symmetry with a
quadratic conserved quantity
$D^{2l}G$\,.

Equations \bref{noether1} and \bref{finalNoether}, on one side,
and \bref{noether2} and \bref{finalNoether2}, on the other, are the two standard
mechanisms for which a Noether symmetry of $L$ is inherited by $\tilde L$.
We will call the corresponding conserved quantities {\sl linear generators} and
{\sl quadratic generators}, respectively.\footnote{The terminology
of {\sl generators} stems
from the canonical framework, and the action of these generators is produced
by way of the Poisson bracket. There is the subtle point, however, that there may exist
Noether symmetries in tangent space that can not be brought --i.e., projected-- to
phase space. In such case the connection of the infinitesimal transformation with
the conserved quantity needs more elaboration (see \cite{Garcia-Pons2000}).}
Let us make some comments on these two mechanisms.

\vspace{6mm}

(\emph{i}) Summarizing the main result of this section,
we observe that the presence of the
classical solution --the background-- causes a
splitting of the original Noether symmetries according to whether they
preserve the background, or are broken by it.
Those that are broken by the background, equation \bref{finalNoether}
(first mechanism),
will yield symmetries for the quadratic fluctuations
action with
{\sl linear} generators.
Instead, the symmetries that preserve
the background, equation \bref{finalNoether2} (second mechanism),
will yield symmetries with
{\sl quadratic} generators.

\vspace{6mm}

(\emph{ii}) Note that gauge symmetries for the fluctuations theory
can only be realized through the first mechanism
\bref{finalNoether}. In fact, since gauge symmetries are generated
in phase space by appropriate combinations of First Class
constraints $\phi_i$, the correspondence between First Class
constraints of both theories already indicates that if
$$
\eta\phi_1 + \dot\eta\phi_2 +...
$$
is a generator of  Noether symmetries
(with the gauge parameter $\eta(t)$) for the original theory,
then
$$
\eta\tilde\phi_1 + \dot\eta\tilde\phi_2 +...
$$
with $\tilde\phi_i = D^l\phi_i$, is a generator of gauge
Noether symmetries for the fluctuations theory, with the
additional fact, already pointed out, that these symmetries
of $\tilde L$ are Abelian.

Noticing that the constraints of the fluctuations theory are linear, we infer that
the gauge transformations for $\tilde L$ must be realized exclusively
through the first mechanism \bref{finalNoether}. A somewhat unexpected
consequence of this fact is the general result that gauge symmetries in the
original theory that completely preserve the background cannot exist,
otherwise the fluctuations theory would change the number of
physical degrees of freedom.
Only for particular restrictions on the gauge parameters,
the gauge symmetries may preserve the background.
Generally covariant theories --having solutions which may exhibit some Killing
symmetries-- and Yang-Mills gauge theories are obvious verifications
of this assertion.

\vspace{6mm}

(\emph{iii}) Since the symmetries provided by
\bref{noether2},
that is, the Noether symmetries of $\tilde L$ inherited from the
background-preserving  symmetries of $L$,
are always rigid, they do not change the number of physical
degrees of freedom of the fluctuations theory. Note that
they do not exist around any solution $q^o$,
but only around particular backgrounds, for which $\delta q|_o=0$.

\vspace{6mm}

(\emph{iv}) We can keep looking at
even higher orders of $G$ in $\epsilon$, generating,
for instance, the transformation law $\tilde{\tilde\delta}Q
=D^{2l}\delta q$. These transformations will emerge as
rigid symmetries of $\tilde L$ when both
$\delta q|_o $ and $D^l\delta q$ vanish. Therefore, the more non-linear
the original theory is, the richer the structure of inherited
rigid symmetries in the fluctuations theory is likely to be.

\vspace{6mm}

(\emph{v}) In this section Lagrangian Noether symmetries have been
studied. We can proceed in a similar way with a Hamiltonian
generator $G^H$ and, starting from \bref{p-expansion}, find the
corresponding linear and quadratic generators in the canonical
fluctuations theory. For transformations projectable to phase
space, and belonging to the type that breaks the background, it
can be shown that if $G^H$ satisfies the conditions (spelled out
in \cite{Batlle-Gomis-Gracia-Pons}) to be a generator of canonical
Noether symmetries for the original theory, then $D^hG^H$
satisfies the same conditions for the fluctuations theory and
becomes a generator of canonical Noether symmetries for it.

\vspace{6mm}

(\emph{vi}) The obstruction to the projectability of Noether
transformations from tangent space to phase space \cite{Garcia-Pons2000}
is related to
the existence of a non-Abelian structure for the --primary and secondary--
First Class constraints of the theory. In consequence,
the symmetries of $\tilde L$ produced by the first mechanism
\bref{noether1} are always projectable to phase space.
In fact, they are field independent transformations.

\vspace{6mm}

(\emph{vii}) One can easily verify that
the projectability to phase space of the rigid Noether symmetries provided by
\bref{noether2} is directly related to the projectablility
of the original transformation. The requirement of projectability of a given
transformation $\delta q$ is, using \bref{gamma},
$$
\gamma_\mu^i \frac{\partial}{\partial \dot q^i} \delta q^j =0\,,
$$
and for the transformations $\tilde \delta Q^j = D^l\delta q^j$ the requirement
is, accordingly,
$$
\gamma_\mu^i|_o \frac{\partial}{\partial \dot Q^i} \delta Q^j =0\,,
$$
where the zero modes of Hessian for $\tilde L$ are
just $\gamma_\mu^i|_o$. However, considering that
$$\frac{\partial}{\partial \dot Q^i} \delta Q^j =
\frac{\partial}{\partial \dot Q^i} D^l\delta q^j =
(\frac{\partial}{\partial \dot q^i} \delta q^j)|_o\,,
$$
we can infer
$$
\gamma_\mu^i \frac{\partial}{\partial \dot q^i} \delta q^j =0 \ \
\Rightarrow (\gamma_\mu^i \frac{\partial}{\partial \dot q^i} \delta q^j)|_o =0 \ \
\Rightarrow \ \ \gamma_\mu^i|_o \frac{\partial}{\partial \dot Q^i}
\delta Q^j =0\,,
$$
which proves our assertion.

\vspace{6mm}

\section{Examples}
\subsection{Massive relativistic free particle \label{freepart}}

Consider the Lagrangian of a massive free particle in Minkowski spacetime,
\beq
L= - m \sqrt{-\dot{\bf x}^2}\,,
\label{L-freepart}
\eeq
with  $\dot{\bf x}^2=\eta _{\mu \nu }\,\dot{x}^\mu \dot{x}^\nu$
and $\dot x^\mu=\frac{dx^\mu}{d\tau }$, where $\eta_{\mu\nu}$ is
the Minkowski metric with \textquotedblleft
mostly plus\textquotedblright\ signature.
There are no Lagrangian constraints even though
the Hessian $W_{\mu\nu} =\frac{m}{\sqrt{-\dot{\bf x}^2}}\,
(\eta_{\mu\nu}-\frac{\dot{x}_\mu \dot{x}_\nu}{{\dot{\bf x}}^2})$ has
one zero mode $\dot x^\nu$ ($W_{\mu \nu }\,\dot x^\nu =0$),
since the Euler-Lagrange e.o.m. are $[L]_{\mu }=-W_{\mu \nu }\,\ddot{x}^\nu $ and
the expression $[L]_\mu \dot{x}^\nu $ vanishes identically.

The Lagrangian has the gauge symmetry of
$\tau$-reparameterizations
$\delta\tau =\varepsilon$, under which the coordinates
transform as $\delta\bf x=\varepsilon\,\bf{\dot x}$ and the Lagrangian
as $\delta L=-\frac{d}{d\tau }\left(\varepsilon L\right)$.
The momentum vector (indices are
raised and lowered with $\eta_{\mu\nu}$)
\bea
\hat{\bf p} =\frac{\partial L}{\partial {\dot{\bf x}}} =
m\,\frac{\dot{\bf x}}{\sqrt{-\dot{\bf x}^2}}\,,
\label{fp-pullback}
\eea
satisfies $\hat{\bf p}^2 + m^2 =0$ identically, thus showing the
existence of a constraint in phase space
\beq
\phi =\frac{1}{2}\,({\bf p}^2 + m^2) \simeq 0\,.
\label{constr-freepart}
\eeq

The canonical Hamiltonian vanishes because the Lagrangian is a
degree one homogeneous function of the velocities. The Dirac
Hamiltonian is therefore
$$
H_D = \lambda\phi\,.
$$
The arbitrary function in phase space $\lambda$ is determined in
tangent space by using the Hamiltonian e.o.m.
$$
\dot{\bf x} = \{{\bf x},\,H_D\}=
\lambda\,\{{\bf x},\,\phi\} = \lambda\bf p\,,
$$
and applying the pullback map $p\rightarrow \hat p$ as defined in
\bref{fp-pullback}. We get
$$\lambda =\frac{\sqrt{-\dot{\bf x}^2}}{m}\,.
$$
$\phi$ is the only constraint in phase space, and
(being the First Class) it generates the gauge transformations
\begin{eqnarray*}
\delta\bf{x} &=& \{\bf{x},\alpha\phi\}=\alpha\bf{p}\,,\\
\delta\bf{p} &=& \{\bf{p},\alpha\phi\}=0\,,
\end{eqnarray*}
with $\alpha(\tau)$ an arbitrary infinitesimal function.
These transformations are $\tau$-repara\-me\-tri\-zations, and they can be put in
the standard Lagrangian form for repara\-me\-tri\-zation invariant theories,
$\delta\bf{x} = \varepsilon \mathbf{\dot x}$, after applying the pullback \bref{fp-pullback} and
redefining the gauge parameter $\alpha (\tau )=\frac{\varepsilon (\tau )}{m}\,
\sqrt{-\mathbf{\dot x}^{2}}$.

Now we will examine the fluctuations theory around a general
solution of the e.o.m. of L. This solution has the form
${\bf x}^o = {\bf u}\,s(\tau) + {\bf c},
$
with ${\bf u},{\bf c}$ constant vectors. We can conventionally
assume that ${\bf u}^2=-1$. $s(\tau)$ is an arbitrary monotonically
increasing function, $\dot s(\tau)=:v(\tau) > 0$.
The fluctuations Lagrangian becomes
$$
\tilde L =\frac{m}{2\, v(\tau)} \,\dot{\bf Q} {\bf \Gamma}\dot{\bf Q}\,,
$$
where ${\bf \Gamma}$ is the projector transversal to the ${\bf u}$ direction,
with the components $\Gamma_{\mu\nu} = \eta_{\mu\nu}+ u_\mu u_\nu$\,.

The canonical Hamiltonian for fluctuations is, according to \bref{H-tilde},
$\tilde H(Q, P)= \bar H(Q, P)+
\lambda(q^o,\dot q^o)B(Q,P)$, where here $\bar H(Q, P)=0$ and
$\lambda(q^o,\dot q^o)=\frac{v(\tau)}{m\,}$. The term $B(Q,P)$ in \bref{phi-expansion}
is now $B(Q,P)= \frac{1}{2}\,{\bf P}^2$, and we obtain
$$
\tilde H(Q, P) =\frac{v(\tau)}{2m}\,{\bf P}^2\,.
$$
One can check that indeed $\tilde H(Q, \hat P)= \hat {\bf P} {\bf Q} -
\tilde L(Q,\dot Q)$.

\subsubsection{Noether symmetries\label{freepartNoether}}
Now we will discuss separately the different symmetries inherited
in the fluctuations theory.

\subparagraph{\emph{Gauge symmetries.}}
The phase space constraint $\tilde \phi$ (recall that $\phi$ expands
as $\phi = \epsilon\tilde \phi + {\cal O}(\epsilon^2)$) becomes
$$\tilde \phi= m ({\bf P}{\bf u})\,,
$$
and it generates the gauge transformations
$\delta {\bf{Q}}=\{{\bf{Q}},\alpha \tilde\phi\}^{\tilde{}} =
\alpha m \bf{u}=\alpha \bf{p}^o$ given by equation \bref{finalNoether}.
\subparagraph{\emph{Rigid symmetries with linear generators.}}
Let us see how the Poincar\'e symmetries of \bref{L-freepart} appear in the
fluctuations theory. The Poincar\'e transformations
$\delta x^\mu = a^\mu+ {\omega^\mu}_\nu x^\nu$, generated by
$G_{p}= p_\mu\,(a^\mu+ {\omega^\mu}_\nu x^\nu)$ (the subindex $p$ is
for Poincar\'e), become
\begin{eqnarray}
\delta Q^\mu&=& \delta x^\mu|_o =  a^\mu+ {\omega^\mu}_\nu ( s(\tau)\,u^\nu +  c^\nu )
= a^\mu+{\omega^\mu}_\nu c^\nu + s(\tau){\omega^\mu}_\nu u^\nu \nonumber \\
&=:& d^\mu + s(\tau)\, r^\mu\,,\label{lorentz-free}
\end{eqnarray}
with $d^\mu$ arbitrary and $r^\mu$ satisfying ${\bf ru}=0$. Both
$d^\mu$ and $r^\mu$ infinitesimal vectors. Since the symmetry is
Abelian and field-independent, the finite transformations have
just the same form with $d^\mu$ and $r^\mu$ finite. The generator
of these symmetries is just $D^hG_p= (d^\mu + s(\tau) r^\mu) P_\mu
+ m r_\mu Q^\mu$. Thus, as Noether symmetries for $\tilde L$ we
obtain the usual translations $\bf d$ and the particular time
dependent translations $s(\tau )\,\bf r$ orthogonal to ${\bf u}$.

The background is preserved for parameters ${\omega^\mu}_\nu$ and $a^\mu$
such that  ${\omega^\mu}_\nu=0$ and $a^\mu+{\omega^\mu}_\nu c^\nu=0$. For this
specific set of parameters, which imply $d^\mu =r^\mu =0$ and hence $D^hG_p=0$,
the Poincar\'e symmetries will be realized with quadratic generators,
as we show later.
\subparagraph{\emph{Other rigid symmetries with linear generators.}}
Besides the gauge and Poincar\'e symmetries, the free particle Lagrangian
in Minkowski spacetime exhibits other Noether symmetries. Take for instance
the quantity $${m\,\bf x\, \Gamma^{(p)} w},$$ with $\bf \Gamma^{(p)}$ being
the projector transversal to the momentum $\bf p$,
$$\Gamma^{({\bf p})}_{\mu\nu} = \eta_{\mu\nu} + \frac{p_\mu p_\nu}{-{\bf p}^2}\,,
$$
and with ${\bf  w}$ being an arbitrary --infinitesimal-- constant
vector. Since this quantity has no explicit time dependence and
has vanishing Poisson bracket with the only constraint, $\phi$, of
the theory, it fulfills the conditions\footnote{Here the notation
\textquotedblleft $\simeq$" means an equality on the primary
constraint surface only.} $\frac{\partial G}{\partial t} + \{G,\,
H\} \simeq\phi$ and $\{G,\, \phi\}  \simeq\phi$ (recall that the
canonical Hamiltonian vanishes in our case), to be a generator of
canonical Noether transformations. In fact we can use a simpler
version for it,
$$ G_g =  {m x^\mu (\eta_{\mu\nu} + \frac{p_\mu p_\nu}{m^2}) w^\nu}\,, $$
where we have used the constraint $\phi$. Now
$$\{G_g,\,\phi\}= \frac{2(\bf wp)}{m}\,\phi\,,$$
which still fulfills the conditions for being a canonical generator.
This generator produces the transformations
$$
\delta {\bf x} = \{ {\bf x},\,G_g\} = \frac{1}{m}\Big(\bf (pw)x + (xp)w\Big)\,,
$$
which, restricted to the background ${\bf x}^o = {\bf u}s(\tau) + {\bf c}$,
${\bf p}^o = m {\bf u}$, give the transformations of the fluctuations
$$
\delta {\bf Q}= \delta {\bf x}|_o =
s(\tau)\Big({\bf -w + (uw)u}\Big) + {\bf (uc)w + (uw)c }\,.
$$
(Note that $\delta {\bf x}|_o=0 \Rightarrow {\bf w}=0\Rightarrow G_g=0$,
thus there is no room in this case for the second mechanism \bref{noether2},
\bref{finalNoether2}.) The piece $s(\tau)\bf (uw)u$ is already included within
the gauge transformations, and the last two pieces are just translations,
already described too. What seems to be a new piece,
$$\delta {\bf Q}=-s(\tau)\bf w\,,$$
is in fact a combination of a gauge transformation in the ${\bf u}$ direction
and the transformation obtained in \bref{lorentz-free}, orthogonal to ${\bf u}$.
\subparagraph{\emph{Rigid symmetries with quadratic generators.}}
The original Lorentz transformations with parameters such that
${\omega^\mu}_\nu u^\nu=0,\ a^\mu+{\omega^\mu}_\nu c^\nu=0$ yield transformations with
quadratic generators for the fluctuations theory. The generator is
$$D^{2h}G_p = P_\mu{\omega^\mu}_\nu Q^\nu$$ with ${\omega^\mu}_\nu u^\nu=0$, where
the operator $D^{2h}$ is defined in phase space as $D^{2l}$ is in tangent space,
shown in equation \bref{D2l-f}. The transformations are
$$\delta Q^\mu= {\omega^\mu}_\nu Q^\nu,$$ again with the rotation parameter
$\omega$ restricted to the
subspace orthogonal to {\bf u}, i.e. with ${\omega^\mu}_\nu u^\nu=0$.
For example, when $u^\mu= \delta_0^\mu$ is
the unit vector along the time direction, the condition
${\omega^\mu}_\nu u^\nu=0$ becomes ${\omega^\mu}_0 =0$, giving the transformations
(in the index notation $\mu = (0, i)$)
$$\delta Q^0=0\,,\qquad\qquad \delta Q^i ={\omega^i}_j\, Q^j,$$
which describe the infinitesimal spatial rotations.


\subsection{Yang-Mills theory}

In a field theory, the coordinates $q^{i}(t)$ are exchanged by the fields
$\phi ^{i,\mathbf{x}}(t):=\phi ^{i}(t,\mathbf{{x})}$, where the spatial point
$\mathbf{x}$ plays the role of a continual index. In consequence, summations
$\sum_{i}$ become integrals $\sum_{i}\int d\mathbf{x}$, derivatives
$\frac{\partial }{\partial q^{i}(t)}$ become functional variations
$\int d\mathbf{{x}\,\frac{\delta }{\delta \phi ^{i}(t,{x})}}$,
while all other variables, for
example momenta $p_{i}(t)$, become densities $\pi _{i}(t,\mathbf{{x})}$. The
Lagrangian density $\mathcal{L}(\phi ,\partial \phi )$ depends on the fields
$\phi $, velocities $\dot{\phi}$ and spatial gradients
$\frac{\partial \phi }{\partial \mathbf{x}}$. The Lagrangian is
then $L(\phi ,\dot{\phi})=\int d\mathbf{{x}\,\mathcal{L}(\phi ,\partial \phi
)}$. All boundary terms are neglected (the fields vanish at the boundary
fast enough) and therefore the Lagrangian density $\mathcal{L}$ is
determined up to a total divergence (and similarly for a Hamiltonian density
$\mathcal{H}$). The basic Poisson bracket is $\left\{ \phi ^{i}\left(
x\right) ,\pi _{j}(x^{\prime })\right\} _{t=t^{\prime }}=\delta
_{j}^{i}\,\delta \left( \mathbf{{x}-{x}^{\prime }}\right) \,,$ but writing the
arguments $x$, $x^{\prime }$ and $\delta $-function will be omitted for the
sake of simplicity.

Consider the Yang-Mills (YM) field theory described by the Lagrangian
(density)
\begin{equation}
\mathcal{L}\left( A,\partial A\right) =-\frac{k}{4}\,F_{\mu \nu
}^{a}F_{a}^{\mu \nu }\,.  \label{YM}
\end{equation}
The gauge field $A_{\mu }^{a}(t,\mathbf{x})=:A_{\mu }^{a}(x)$ and the
associated field strength $F_{\mu \nu }^{a}=\partial _{\mu }A_{\nu
}^{a}-\partial _{\nu }A_{\mu }^{a}+f_{bc}^{a}\,A_{\mu }^{b}A_{\nu }^{c}$
depend on the coordinates $x^{\mu }=(x^{0},x^{i})=(t,\mathbf{{x})}$ of
Minkowski spacetime. The constant $k$ (usually denoted by $1/g_{YM}^{2}$) is
dimensionless and positive. The indices $a,b,\ldots $ label the Lie
generators of a non-Abelian (semi-simple) Lie group with the structure
constants $f_{abc}\ $and the Cartan metric $g_{ab}$.\footnote{
Minkowski metric $\eta _{\mu \nu }$ raises and
lowers spacetime indices, and the Cartan metric $g_{ab}$
raises and lowers group indices.} The YM Lagrangian is invariant under the
gauge transformations $\delta A_{\mu }^{a}(x)=D_{\mu }\alpha ^{a}(x)\equiv
\partial _{\mu }\alpha ^{a}+f_{bc}^{a}\,A_{\mu }^{b}\alpha ^{c}$. The
Euler-Lagrange e.o.m.
\begin{equation}
\lbrack \mathcal{L}]_{a}^{\mu }:=\frac{\partial \mathcal{L}}{\partial A_{\mu
}^{a}}-\partial _{\nu }\frac{\partial \mathcal{L}}{\partial (\partial _{\nu
}A_{\mu }^{a})}=k\,D_{\nu }F_{a}^{\nu \mu }  \label{YM eom}
\end{equation}
have singular Hessian $W_{ab}^{\mu \nu }=k\,g_{ab}\left( \eta ^{\mu \nu
}+\eta ^{\mu 0}\eta ^{\nu 0}\right) $, with zero modes
$(\gamma _{b})_{\mu }^{a}=\delta _{\mu }^{0}\delta
_{b}^{a}\,$, leading to the primary Lagrangian constraints
$[\mathcal{L}]_{a}^{0}=k\,D_{i}F_{b}^{i0}\simeq 0$.

Defining the canonical momenta
\begin{equation}
\hat{\pi}_{a}^{\mu }(A,\partial A)=\frac{\partial \mathcal{L}}{\partial
\dot{A}_{\mu }^{a}}=- k\,F^{0\mu }_{a}\,,  \label{YM pi}
\end{equation}
we pass to Hamiltonian formalism and
find primary ($\pi _{a}^{0}$) and secondary ($\theta _{a}$) constraints
\begin{equation}
\pi _{a}^{0}\simeq 0\,,\qquad \qquad \theta _{a}\equiv
\dot{\pi}_{a}^{0}=D_{i}\pi _{a}^{i}\simeq 0\,,  \label{YM constr}
\end{equation}
with the Dirac Hamiltonian density
\begin{equation}
\mathcal{H}_{D}=\frac{1}{2k}\,\pi _{a}^{i}\pi _{i}^{a}+
\frac{k}{4}\,F_{ij}^{a}\,F_{a}^{ij}-A_{0}^{a}\,
\theta _{a}+\lambda ^{a}\pi _{a}^{0}\,.
\label{YM H_D}
\end{equation}
Lagrange multipliers $\lambda ^a (x)$ can be
determined in tangent space (see Appendix \ref{multipliers}) from
Hamilton-Dirac's equations, $\lambda ^a(A,\dot{A})=\dot{A}^a$.
The components $A_{0}^{a}$ play the role of Lagrange multipliers
for secondary constraints $\theta _{a}$. These constraints do not evolve in
time since $\dot{\theta}_{a}=-f_{ab}^{c}\,A_{0}^{b}\,\theta _{c}\simeq 0$, and
thus there are no new constraints. (In calculation, the identity
$[D_{i},D_{j}]v^{a}=f_{bc}^{a}\,F_{ij}^{b}\,v^{c}$ was used.)
All constraints are First Class and the only nontrivial brackets close
non-Abelian algebra
\begin{equation}
\{\theta _{a},\theta _{b}\}=f_{ab}^{c}\,\theta _{c}\,.  \label{YM algebra}
\end{equation}
The canonical generator
\begin{equation}
G[\alpha ]=\int d\mathbf{x}\,\left( D_{0}\alpha ^{a}\pi _{a}^{0}-\alpha
^{a}\theta _{a}\right)  \label{YM G}
\end{equation}
induces gauge transformations
\begin{equation}
\delta A_{\mu }^{a}=\{A_{\mu }^{a},G[\alpha ]\}=D_{\mu }\alpha ^{a}\,,\qquad
\qquad \delta \pi _{a}^{\mu }=\{\pi _{a}^{\mu },G[\alpha
]\}=-f_{ab}^{c}\,\alpha ^{b}\pi _{c}^{\mu }\,.
\end{equation}
The transformation law for $\pi _{\mu }^{a}$ is consistent with $\delta
\hat{\pi}_{\mu }^{a}$ obtained from (\ref{YM pi}).

\subsubsection{Fluctuations theory}

Consider now the small fluctuations $Q$ expanded as $A=\bar{A}+\epsilon Q$
around a solution $\bar{A}$ of the e.o.m. (\ref{YM eom}). Then the corresponding
field strength expands as
$$
F_{\mu \nu }^{a}=\bar{F}_{\mu \nu }^{a}+\epsilon \left( \bar{D}_{\mu }Q_{\nu
}^{a}-\bar{D}_{\nu }Q_{\mu }^{a}\right) +\epsilon ^{2}f_{bc}^{a}\,Q_{\mu
}^{b}Q_{\nu }^{c}\,,
$$
where all hatted operators denote these operators evaluated at
$\bar{A}$. Therefore, we find the Lagrangian for fluctuations theory
\begin{equation}
\tilde{\mathcal{L}}(Q,\partial Q)=-\frac{k}{2}\left[ \bar{D}^{\mu
}Q_{a}^{\nu }\left( \bar{D}_{\mu }Q_{\nu }^{a}-\bar{D}_{\nu }Q_{\mu
}^{a}\right) +f_{bc}^{a}\,\bar{F}_{a}^{\mu \nu }Q_{\mu }^{b}Q_{\nu }^{c}
\right] \,.  \label{YM eff}
\end{equation}
The canonical formulation of (\ref{YM eff}) can be obtained directly from
the original Hamiltonian analysis, by expanding it as $A=\bar{A}+\epsilon Q$
and $\pi =\bar{\pi}+\epsilon P$ around a solution $\bar{A}$, $\bar{\pi}$
of the canonical e.o.m. Here the basic bracket is $\{Q,P\}^{\tilde{}}=1$.
Linear terms of the original constraints (\ref{YM constr}) give the
primary ($P_{a}^{0}$) and secondary ($\tilde{\theta}_{a}$) constraints in
the fluctuations theory
$$
P_{a}^{0}\simeq 0\,,\qquad \qquad \tilde{\theta}_{a}=
\bar{D}_{i}P_{a}^{i}+\,f_{ab}^{c}\,Q^{b}_{i}\,\bar{\pi}_{c}^{i}
\simeq 0\,.
$$
The Dirac Hamiltonian (\ref{YM H_D}) with the multipliers $\dot{A}_{0}^{a}=
\frac{d\bar{A}_{\mu }^{a}}{dt}+\epsilon \dot{Q}_{0}^{a}\equiv \,
\bar{\lambda}
+\epsilon \tilde{\lambda}$ expands as
$$
\mathcal{H}_{D}\left( A,\pi ,\lambda \right) =
{\mathcal{H}}(\bar{A}_{\mu }^{a},\bar{\pi}_{a}^{\mu })+\epsilon \,
(\frac{d\bar{A}_{\mu }^{a}}{dt}\,P_{a}^{\mu }
-\frac{d\bar{\pi}_{a}^{\mu }}{dt})\,Q_{\mu }^{a}+\epsilon ^{2}\tilde{\mathcal{H}}_{D}
(Q,P,\tilde{\lambda})+
\mathcal{O}(\epsilon ^{3})\,,
$$
where $\tilde{\mathcal{H}}_{D}=\tilde{\mathcal{H}}+\tilde{\lambda}
^{a}P_{a}^{0}$. The canonical Hamiltonians of the fluctuations theory is
$$
\tilde{\mathcal{H}}=\frac{1}{2k}\,P_{a}^{i}P_{i}^{a}+\frac{k}{2}\,
\bar{D}^{i}Q_{a}^{j}\left( \bar{D}_{i}Q_{j}^{a}-\bar{D}_{j}Q_{i}^{a}\right)
+\frac{k}{2}\,f_{bc}^{a}\,\bar{F}_{a}^{ij}Q_{i}^{b}Q_{j}^{c}+f_{bc}^{a}\,
\bar{A}_{0a}P_{b}^{j}Q_{j}^{c}-Q_{0}^{a}\,\tilde{\theta}_{a}\,.
$$
This result can also be obtained directly from $\tilde{\mathcal{H}}=\hat{P}Q-%
\mathcal{L}$. We find that the constraint algebra is Abelianized,
$$
\{\tilde{\theta}_{a},\tilde{\theta}_{b}\}^{\tilde{}}=f_{ab}^{c}\,
\bar{\theta}_{c}\equiv 0\,,
$$
as expected from \bref{equalbrackets}.

Consider now symmetries of the
fluctuations theory. The gauge generator (\ref{YM G}) expands as $G[\alpha
]=\epsilon \tilde{G}[\alpha ]+\epsilon \,\widetilde{\widetilde{G}}[\alpha ]$,
where the linear ($\tilde{G}$) and quadratic ($\widetilde{\tilde{G}}$)
generators are
\bea
\tilde{G}[\alpha ] &=&\int d{\bf x}\,(\bar{D}_{0}\alpha
^{a}P_{a}^{0}-\alpha ^{a}\tilde{\theta}_{a})\,, \\
\widetilde{\widetilde{G}}[\alpha ] &=&-\int d\mathbf{x}\,f_{ab}^{c}\,\alpha
^{a}Q_{\mu }^{b}P_{c}^{\mu }.
\eea
Linear generator induces Abelianized gauge symmetries
$\delta Q_{\mu }^{a}=\bar{D}_{\mu }\alpha ^{a}$,
in agreement with \bref{finalNoether}.
$\widetilde{\widetilde{G}}[\alpha ]$ is a generator of rigid symmetries in
fluctuations theory only if there are such parameters $\alpha ^{a}$ and
vacuums $\bar{A}^{a}$, for which $\delta \bar{A}_{\mu }^{a}=\bar{D}_{\mu
}\alpha ^{a}=0$. Then the transformations
$$
\tilde{\delta}Q_{\mu }^{a}=f_{bc}^{a}\,Q_{\mu }^{b}\alpha ^{c}\,,\qquad
\qquad \tilde{\delta}P_{a}^{\mu }=-f_{ab}^{c}\,P_{c}^{\mu }\alpha ^{b}\,,
$$
leave $\tilde{\mathcal{H}}$ invariant. Similarly, $\tilde{\delta}Q$ leaves
$\tilde{\mathcal{L}}$ invariant. Indeed, under the transformations
$\tilde{\delta}Q_{\mu}^{a}=f_{bc}^{a}Q_{\mu }^{b}\alpha ^{c}$,
the Lagrangian \bref{YM eff} changes as
\[
\tilde{\delta}\tilde{\mathcal{L}}=-k\,f_{abc}\left[ \left( \bar{D}_{\mu
}Q_{\nu }^{a}-\bar{D}_{\nu }Q_{\mu }^{a}\right) Q^{\nu b}+Q_{\mu
}^{a}\,Q_{\nu }^{b}\,\bar{D}^{\nu }\right] \bar{D}^{\mu }\alpha ^{c}\,,
\]
where the identity $f_{abc}\bar{D}_{[\mu }Q_{\nu ]}^{a}\bar{D}^{[\mu
}Q^{\nu ]b}\equiv 0$, the Jacobi identity
$f_{a[b}^{c}\,f_{de]c}\equiv 0$ and $[\bar{D}_{\mu },\bar{D}_{\nu }]
\alpha ^{c}=f_{ae}^{c}\bar{F}_{\mu _{\nu }}^{a}\alpha ^{e}$
have been used.\footnote{$[\cdots ]$
acts on the indices inside the bracket as
the antisymmetrization.} In that way, for the backgrounds for which
$\bar{D}\alpha ^{a}=0$, the fluctuations Lagrangian becomes invariant,
$\tilde{\delta}\tilde{\mathcal{L}}=0$.
The condition $\bar{D}\alpha^a=0$ on the gauge parameter
$\alpha^a$ is not trivial.
The existence of a solution
depends on the topology of a manifold where the YM theory is defined
and on the boundary conditions for $\alpha^a$, as well as on the properties
of the background $\bar{A}$ (such as its possible winding numbers, etc.).
Additionally, $\alpha^a$ has to be globally defined.

\section{Conclusions}

This work is dedicated to the study of the dynamics of small
fluctuations oscillating around a classical solution of a gauge
theory. This system, where the fluctuation is a fundamental field,
is described by the explicitly time-dependent quadratic Lagrangian
or Hamiltonian. We show that in First Order systems, that is,
ordinary tangent space or phase space formulations, it is
permitted to choose freely between the Lagrangian and Hamiltonian
formalism, with the certainty that the same result will be reached
at the end, assuming that the conditions 
(\emph{i})-(\emph{iii}) at the begining of Section 3, have
been fulfilled. We show that in such a case, the Legendre
transformation commutes with the transformation which maps the
original Lagrangian to the quadratic one, at first order in the
fluctuations expansion. In fact, the mismatch in Lagrangian and
Hamiltonian descriptions occurs in higher orders in the expansion
parameter $\epsilon$, but it does not affect the consistency of
neither of these two descriptions.

Other results of our analysis are the following. While the
fluctuations Lagrangian is defined as the quadratic term of the
original Lagrangian, the fluctuations canonical Hamiltonian
contains, apart from the quadratic term coming from the original
canonical Hamiltonian, also the contribution of the quadratic part
of primary constraints and the corresponding Lagrange multipliers
computed for the classical solution under consideration.

Furthermore, we prove that, under the assumptions made in Section
3, this mapping, or \textquotedblleft
linearization\textquotedblright\ of the theory (since the
equations of motion become linearized by it), entirely keeps the
structure of the original theory in the fluctuations theory as
well. For example, the class of constraints (First or Second) does
not change after the linearization, and the structure of the
constraint algorithm remains the same. Since the linearization
does not change the number of First and Second class constraints,
it follows that the number of physical degrees of freedom in both
theories is the same.

We address in particular the issue of Noether symmetries. As
regards the gauge ones, we find that the First Class constraints
in the fluctuations theory always correspond to Abelian gauge
symmetries, since the non-Abelian symmetry cannot be described by
quadratic Lagrangians. Another interesting outcome is that the
choice of the background may influence the expression of the rigid
Noether symmetries in the fluctuations theory. One part of these
symmetries (the one which mimics those of the original theory) is
generated by the linear terms of the fluctuations expansion of the
original generators. If, however, it happens that some background
is preserved by a subset of the original rigid symmetries, then
the fluctuations theory exhibits rigid symmetries coming from the
quadratic powers in the fluctuations of the original generator. In
supersymmetric theories this is the way for instance in which the
symmetries preserved by a BPS state are realized in the
fluctuations theory.

As mentioned above, our results are reliable for the
systems with constant Hessian and around non-de\-ge\-ne\-ra\-te
solutions. For the systems with degenerate solutions, the linear
approximation is not applicable any longer. For example, a
degenerate solution $q^{o}$ can lead to the ineffective
constraints (of the type $(q-q^{o})^{2}\simeq 0$). After the
linearization, these constraints vanish, what effectively leads to
the increase in the number of degrees of freedom. One example of a
such degenerate background in Chern-Simons supergravity is
presented in ref. \cite{Chandia-Troncoso-Zanelli} and
they are treated in Hamiltonian formalism in ref.
\cite{Miskovic-Troncoso-Zanelli}. In Lagrangian formalism, they
have not been studied yet.


\section*{Acknowledgments}

O. M. would like to thank Jorge Zanelli for helpfull discussions.
J. M. P. thanks Joaquim Gomis and Jorge Russo for discussions and
useful comments. This work was partially funded by Chilean
FONDECYT grant 3040026, the European Community's Human Potential
Programme under contract MRTN-CT-2004-005104 `Constituents,
fundamental forces and symmetries of the universe', the Spanish
grant MYCY FPA 2004-04582-C02-01, and by the Catalan grant CIRIT
GC 2001, SGR-00065.


\appendix
\section{Identities with the bracket expansions \label{expansions}}

Here we give some results used in the main text. For generic
expansions of functions $f(q,p;t)$, $g(q,p;t)$,
\begin{eqnarray}
f(q,p;t)&=& f_o(t) + \epsilon f_1(Q,P;t)
+ \epsilon^2 f_2(Q,P;t) + {\cal O}(\epsilon^3)  \nonumber \\
g(q,p;t)&=& g_o(t) + \epsilon g_1(Q,P;t)
+ \epsilon^2 g_2(Q,P;t)+ {\cal O}(\epsilon^3)\,,\label{expand-f-g}
\end{eqnarray}
(where $f_o(t) =f(q^o(t),p^o(t);t),\ g_o(t)=g(q^o(t),p^o(t);t)$ and $f_1=D^h f,\,g_1=D^h g $)
the following relations hold
\bea
\{f_1,\,g_1\}^{\tilde{}}= \{f,\,g\}\!|_o\,, \label{ap1}
\eea
\bea
D^h\{f,\,g\}= \{f_2,\,g_1\}^{\tilde{}} + \{f_1,\,g_2\}^{\tilde{}}\,.\label{ap2}
\eea
The proofs are immediate, using \bref{expand-f-g} and the
corresponding expansion for the bracket $\{f,\,g\}$.

\vspace{6mm}

Explicit time dependence has a different meaning in the original theory and in the
fluctuations theory because the change of variables
$q\rightarrow q^o + \epsilon Q$ is time dependent ---$q^o$ is in fact the trajectory $q^o(t)$.
In particular we have,
in the canonical formalism (similar results hold in the tangent space formulation),
\begin{eqnarray}
\frac{\partial^{{}^f}}{\partial Q} &=&\epsilon\,\frac{\partial}{\partial q}\nonumber \\
\frac{\partial^{{}^f} }{\partial P} &=&\epsilon\,\frac{\partial}{\partial p}\nonumber \\
\frac{\partial^{{}^f} }{\partial t} &=&\frac{\partial }{\partial t} +
\dot q^o\frac{\partial}{\partial q}+\dot p^o\frac{\partial}{\partial p}\,,
\end{eqnarray}
where the superscript ${}^f$ stands for the fluctuation variables and is conventionally
omitted in the text. Note that
$$
 \frac{d}{d t}= \frac{\partial }{\partial t}+
 \dot q\frac{\partial}{\partial q}
+ \dot p\frac{\partial}{\partial p}+...=\frac{\partial^{{}^f}  }{\partial t}+
\dot Q\frac{\partial^{{}^f} }{\partial Q}+ \dot P\frac{\partial^{{}^f} }{\partial P}+...\,,
$$
(where the dots indicate higher-order tangent structures like
$\ddot q\frac{\partial}{\partial\dot q}
+ \ddot p\frac{\partial}{\partial\dot p}$ etc.) as it must be.

One can also derive the following result used in the text,
\bea
[D^h,\,\frac{\partial }{\partial t}]f = \{P \dot q^o - Q \dot
p^o,\, f_2\}^{\tilde{}}\,. \label{ap3} \eea Obviously the l.h.s.
must be understood as $D^h\circ\frac{\partial }{\partial t}-
\frac{\partial^{{}^f} }{\partial t}\circ D^h.$ The proof of
\bref{ap3} is
\begin{eqnarray}
[D^h,\,\frac{\partial }{\partial t}]f &=& Q \frac{\partial f}{\partial q \partial t}|_o
+ P \frac{\partial f}{\partial p \partial t}|_o - \frac{\partial }{\partial t}(
Q \frac{\partial f}{\partial q }|_o
+ P \frac{\partial f}{\partial p }|_o)\nonumber \\
&=&- \Big(Q \frac{\partial f}{\partial q \partial q}|_o \dot q^o +
Q \frac{\partial f}{\partial q \partial p}|_o \dot p^o +
P \frac{\partial f}{\partial p \partial q}|_o \dot q^o  +
P \frac{\partial f}{\partial p \partial p}|_o \dot p^o \Big)\nonumber \\
&=& - \dot q^o\frac{\partial f_2}{\partial Q}-\dot p^o\frac{\partial f_2}{\partial P}
=\{P \dot q^o - Q \dot p^o,\, f_2\}^{\tilde{}}\,.
\end{eqnarray}
\section{Multipliers in Hamilton-Dirac equations of motion}
\label{multipliers}
The Hamilton-Dirac e.o.m. are
\begin{eqnarray}
\dot q &=& \frac{\partial H}{\partial p} +
\lambda^\mu\frac{\partial \phi_\mu}{\partial p}\,, \nonumber \\
\dot p &=& -\frac{\partial H}{\partial q}
-\lambda^\mu\frac{\partial \phi_\mu}{\partial q}\,,\nonumber \\\label{eomH2}
 0 &=&\phi^{(0)}_\mu (q,p)\,,
\end{eqnarray}
where the $\lambda^\mu$ are in principle arbitrary functions of time
(in fact they may also be arbitrary functions of the $q$ and $p$ variables
in the Hamiltonian approach,
but this is not relevant for our discussion).
Not any choice for these functions is allowed in the formalism, as it is seen
when one performs the Dirac-Bergman constraint algorithm,
where eventually some of these
functions become determined in phase space whereas some others stay completely
arbitrary and in fact describe the gauge freedom
(local symmetries) contained in the dynamics.

In spite of their initial arbitrariness in phase space, it is
interesting to notice that one can always determine these
arbitrary functions $\lambda^\mu$ as definite functions in tangent
space. To this end one must first use the first equation in
\bref{eomH2} and apply the pullback $p\rightarrow \hat p$; then,
since the matrix $\frac{\partial \phi_\mu}{\partial p}$ has
maximum rank, the algebraic equation for the $\lambda^\mu$'s, \beq
\dot q = {\mathcal F}\!L^*\frac{\partial H}{\partial p} +
\lambda^\mu{\mathcal F}\!L^*\frac{\partial \phi_\mu}{\partial
p}\,, \label{lambda-determ} \eeq can be solved for all
$\lambda^\mu$ as functions $\lambda_{def}^\mu(q,\dot q)$ defined
in tangent space. The rationale for this construction is as
follows: if for some given set of arbitrary functions
$\lambda^\mu(t)$ we obtain a solution $q(t), p(t)$ of
\bref{eomH2},
then $\lambda_{def}^\mu(q(t),\dot q(t)) =
\lambda^\mu(t)$. In fact the e.o.m. \bref{eomH2} are completely
equivalent to the following e.o.m.
\begin{eqnarray}
\dot q &=& \frac{\partial H}{\partial p} +
\lambda_{def}^\mu(q,\dot q)\frac{\partial \phi_\mu}{\partial p}\,, \nonumber \\
\dot p &=& -\frac{\partial H}{\partial q}
-\lambda_{def}^\mu(q,\dot q)\frac{\partial \phi_\mu}{\partial q}\,,
\label{eomH3}\nonumber \\
0 &=&\phi^{(0)}_\mu (q,p)\,.
\end{eqnarray}
Thus one can either work with arbitrary functions $\lambda^\mu(t)$, as in \bref{eomH2},
or just consider that the e.o.m. are \bref{eomH3}. In \bref{eomH3} we see that the unknown
$\dot q$ appears not only in the l.h.s., but also in the r.h.s.
The impossibility to write \bref{eomH3} in normal form\footnote
{Differential equations are written in normal form
when the highest derivatives can be
isolated in the l.h.s.} is signaling the possible presence of
gauge freedom.



\end{document}